\documentclass[9pt]{article}

\usepackage{amsmath}
\usepackage{amssymb}
\usepackage{amsfonts}
\usepackage{verbatim}
\usepackage{graphicx}
\usepackage{palatino,url,multicol}
\usepackage{cite}

\usepackage{float}
\usepackage{caption}
\usepackage{subcaption}

\usepackage[margin=1in]{geometry}

\graphicspath{{converted_graphics/}}

\begin{document}

\title{Delay Effects on Amplitude Death, Oscillation Death, and Renewed Limit Cycle Behavior in Cyclically Coupled Oscillators}

\author{Ryan Roopnarain and S. Roy $\text{Choudhury}^*$\\
Department of Mathematics\\
University of Central Florida\\
sudipto.choudhury@ucf.edu }
\date{\today}

\maketitle

\begin{abstract}

\textit{The effects of a distributed 'weak generic kernel' delay on cyclically coupled limit cycle and chaotic oscillators are considered. For coupled Van der Pol oscillators (and in fact, other oscillators as well) the delay can produce transitions from amplitude death(AD) or oscillation death (OD) to Hopf bifurcation-induced periodic behavior, with the delayed limit cycle shrinking or growing as the delay is varied towards or away from the bifurcation point respectively. The transition from AD to OD is mediated here via a pitchfork bifurcation, as seen earlier for other couplings as well. Also, the cyclically coupled undelayed van der Pol system here is already in a state of AD/OD, and introducing the delay allows both oscillations and AD/OD as the delay parameter is varied. This is in contrast to other limit cycle systems, where diffusive coupling alone does not result in the onset of AD/OD.}

\textit{For systems where the individual oscillators are chaotic, such as a Sprott oscillator system or a coupled van der Pol-Rayleigh system with parametric forcing, the delay may produce AD/OD (as in the Sprott case), with the AD to OD transition now occurring via a transcritical bifurcation instead. However, this may not be possible, and the delay might just vary the attractor shape. In either of these situations however,  increased delay strength tends to cause the system to have simpler behavior, streamlining the shape of the attractor, or shrinking it in cases with oscillations.}


\end{abstract}

\section{Introduction}

Cooperative behaviors in coupled oscillators have been actively studied in various fields in recent years\cite{Str}. Various such phenomena include several kinds of synchronization\cite{Boc}, quenching of oscillations, phase locking, and complex chimera states\cite{Pik}. 

Oscillation quenching\cite{Sax, Kos} has applications in a variety of biological and chemical systems\cite{Ull}-\cite{Yos}, and may occur via a variety of couplings, as well as via both discrete and distributed time delays of sufficient strength\cite{Bar}-\cite{Hen}.

Quenched states are now distinguished into two categories, viz. amplitude death (AD) and oscillation death (OD). The former (AD) occurs when all the coupled sub-systems settle to a common stable and homogeneous steady state (HSS) or fixed point. By contrast, the latter (OD) corresponds to the various oscillators settling to or populating different, coupling-dependent stable states, referred to as inhomogeneous steady states (IHSS). In some systems, coexistence of HSS and IHSS behaviors\cite{Kos1}, or of OD with limit cycles\cite{Ruw}, or multi-cluster OD and other more complex states\cite{Rak} in networks, or bifurcations of limit cycles to more complex oscillatory states\cite{Ban} have also been observed.

Transitions from HSS to IHSS states are of significant interest in physical phenomena, a classical example being the diffusion induced Turing instability\cite{Tur} leading to the formation of pattern from a homogeneous background. For instance, such behavior has been observed\cite{Sax} in systems with diffusive coupling, discrete delay, conjugate coupling, dynamic coupling, repulsive interaction, mean-field coupling, and linear augmentation\cite{Kos2}-\cite{Sha}. If the individual oscillators are of the limit cycle variety, the symmetry breaking from the HSS to the IHSS state has generally been found to occur via a pitchfork bifurcation, irrespective of the coupling or other symmetry breaking features of the system. In chaotic oscillators, the situation is more complicated, and that will be one of our primary areas of focus in this paper.

The remainder of the paper is organized as follows. Section 2 considers the linear stability analysis, and local bifurcations of a system of Van der Pol oscillators, and a chaotic Sprott system, both cyclically coupled and with a distributed delay incorporated. Section 3 considers detailed numerical results for both systems, including various parameter regimes and types of dynamics. The results and conclusions are summarized in Section 4

\section{Linear Stability and Local Bifurcation Analysis of Cyclically Coupled and Delayed Systems}

In this section we consider the linear stability of the cyclically coupled and delayed limit cycle and chaotic systems which we will be considering. 


\subsection{Van Der Pol Oscillators with Cyclic Coupling and Delay}
First consider a system of Van Der Pol Equations under cyclic coupling given by
\begin{align}
\label{undelayvdp}
\dot{x}_1 &= \omega_1 y_1 + \varepsilon_1 (x_2 - x_1) \notag \\
\dot{y}_1 &= b (1 - x_1^2) y_1 - \omega_1  x_1 \notag\\
\dot{x}_2 &= \omega_2 y_2 \notag\\
\dot{y}_2 &= b (1 - x_2^2) y_2 - \omega_2 x_2 + \varepsilon_2 (y_1 - y_2)
\end{align}
where $\varepsilon_{1,2}$ are the coupling strengths, $\omega_{1,2}$ are the frequencies, and we take $b = 3/10$ \cite{Olu}.

Introducing a weak distributed time delay in the last equation:
\begin{align}
\label{vdpdelay1}
\dot{x}_1 &= \omega_1 y_1 + \varepsilon_1 (x_2 - x_1) \notag \\
\dot{y}_1 &= b (1 - x_1^2) y_1 - \omega_1  x_1 \notag\\
\dot{x}_2 &= \omega_2 y_2 \notag\\
\dot{y}_2 &= b (1 - x_2^2) y_2 - \omega_2 x_2 + \varepsilon_2 \left(\int_{-\infty}^t ay_1(\tau)e^{-a(t-\tau)}d\tau  - y_2\right)
\end{align}
and defining 
\begin{equation}
z(t) = \int_{-\infty}^t ay_1(\tau)e^{-a(t-\tau)}d\tau\notag
\end{equation}
we can reduce the system \eqref{vdpdelay1} to the system of ordinary differential equations:
\begin{align}
\label{vdpdelay}
\dot{x}_1 &= \omega_1 y_1 + \varepsilon_1 (x_2 - x_1) \notag \\
\dot{y}_1 &= b (1 - x_1^2) y_1 - \omega_1  x_1 \notag\\
\dot{x}_2 &= \omega_2 y_2 \notag\\
\dot{y}_2 &= b (1 - x_2^2) y_2 - \omega_2 x_2 + \varepsilon_2 \left(z - y_2\right)\notag\\
\dot{z} &= a(y_1-z)
\end{align}

The fixed points of the delayed system are the trivial fixed point $P_0$: 
\begin{equation}
P_0 = (x_1,y_1,x_2,y_2) = (0, 0, 0, 0, 0)
\end{equation}
and two nontrival fixed points given by:
\begin{align}
   P_1 &=  (x_+, y_+, \varepsilon_2 y_+/\omega_2,0, y_+) \\
   P_2 &= (x_-, y_-, \varepsilon_2 y_-/\omega_2, 0, y_-)
\end{align}
where
\begin{equation}
    x_{\pm} = \pm \sqrt{1-\frac{\omega_1^2\omega_2 + \varepsilon_1\varepsilon_2\omega_1}{b\omega_2\varepsilon_1}}
\end{equation}
\begin{equation}
    y_{\pm} = \frac{\omega_1 x_{\pm}}{b(1-x_{\pm}^2)}
\end{equation}
In this paper we will consider the case where $\varepsilon_1 = \varepsilon_2 = \varepsilon$. Following the methods of phase-plane analysis, the eigenvalues of the Jacobian matrix of \eqref{vdpdelay} 
evaluated at the fixed point $P_0$ (and with $b=3/10$) satisfy the characteristic equation
\begin{align} 
\label{delayvdppoly-trivial}
&\lambda^5+  \left(a - 2b + 2 \varepsilon \right)\lambda^4 + \left(a \left(2 \varepsilon-\frac{3}{5}\right)+\varepsilon^2-\frac{9 \varepsilon}{10}+\omega_1^2+\omega_2^2+\frac{9}{100}\right) \lambda^3\notag \\
&\quad + \left(a \left(\varepsilon^2-\frac{9 \varepsilon}{10}+\omega_1^2+\omega_2^2+\frac{9}{100}\right)-\frac{3 \varepsilon^2}{10}+\varepsilon \left(\omega_1^2+\omega_2^2+\frac{9}{100}\right)\right.\notag\\
&\quad \left.-\frac{3}{10} \left(\omega_1^2+\omega_2^2\right)\right) \lambda^2 +\left(\frac{1}{100} a \left(-30 \varepsilon^2+\varepsilon \left(100 \omega_1^2+100 \omega_2^2+9\right)\right.\right.\notag\\
&\quad\left.\left.-30 \left(\omega_1^2+\omega_2^2\right)\right)-\frac{3 \varepsilon \omega_2^2}{10}+\omega_1^2 \omega_2^2 \right)\lambda + \frac{1}{10} a \omega_2 \left(10 \varepsilon^2 \omega_1-3 \varepsilon \omega_2+10 \omega_1^2 \omega_2\right)=0
\end{align}

Similarly, and also setting $b = 3/10$, the eigenvalues of the Jacobian matrix of \eqref{vdpdelay} at either of the nontrivial fixed points $P_1$ or $P_2$ satisfy the (same) characteristic equation:

\begin{equation} \label{characnew}
\lambda^5 + b_1\lambda^4 + b_2\lambda^3+b_3\lambda^2+b_4\lambda+b_5=0
\end{equation}
where the coefficients $b_i, i = 1, 5$ are given in Appendix A.

For the fixed point $P_i$, $i=0,1,2$, to be a stable fixed point within the linearized analysis, all the associated eigenvalues must have negative real parts. From the Routh-Hurwitz criteria, the necessary and sufficient conditions for a fifth degree polynomial equation of the form:

\begin{equation}\label{generalcharpoly5}
\lambda^5 + b_1\lambda^4 + b_2\lambda^3 + b_3\lambda^2 + b_4\lambda+b_5 = 0
\end{equation}

\noindent
to have $\hbox{Re} (\lambda_{1,2,3,4,5}) <0$ are:
\begin{align}
\label{vrh1}
b_1 &>0 \\ 
\label{vrh5}
b_5 &>0 \\ 
\label{vrh2}
b_1 b_2 - b_3 &>0 \\ 
\label{vrh3}
b_1 (b_2 b_3 + b_5) -b_3^2 - b_1^2 b_4 &>0 \\
\label{vhopfcondition}
b_1 (b_2 b_3 b_4 - b_2^2 b_5 + 2 b_4 b_5)-b_3^2 b_4- b_5^2 + b_2b_3b_5-b_1^2b_4^2 &>0 
\end{align}

\begin{figure}
\centering
\centerline{\includegraphics[width=0.55\textwidth]{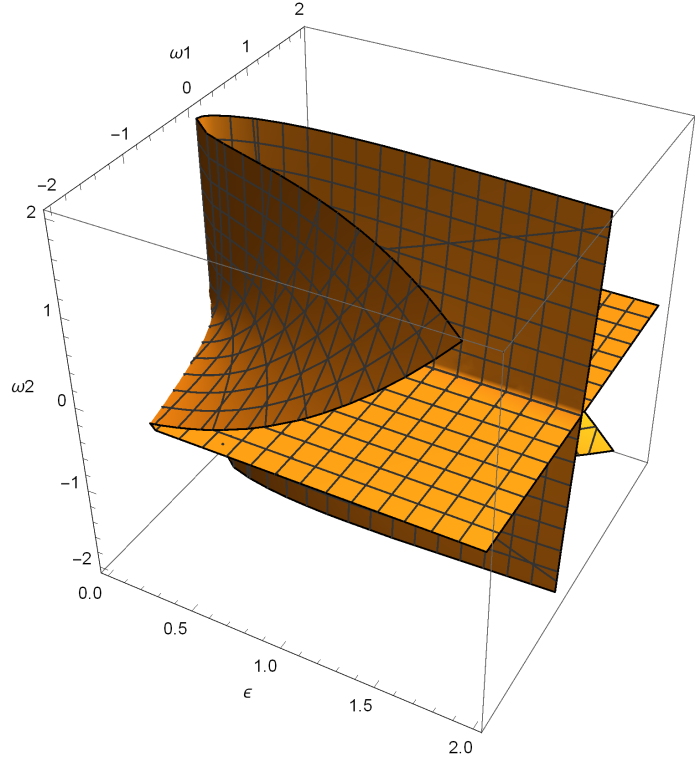}}
\caption{\label{fig:pitchfork} Pitchfork bifurcation surface of trivial fixed point of \eqref{vdpdelay1} for $b = 3/10$ and $\varepsilon_1 = \varepsilon_2 = \varepsilon$.}
\end{figure}

It is straightforward to check that, for general values of $b$, the fixed point $P_0$ undergoes a supercritical pitchfork bifurcation when
\begin{equation} \label{pitchfork}
1-\frac{\omega_1^2\omega_2 + \varepsilon_1\varepsilon_2\omega_1}{b\omega_2\varepsilon_1} = 0
\end{equation}
with $P_0$ going unstable, and the two non-trivial fixed points being born (and being stable) when the expression on the left becomes positive. For $b = 3/10$ and $\varepsilon_1 = \varepsilon_2 = \varepsilon$, this pitchfork bifurcation surface is plotted in Figure \ref{fig:pitchfork}.

When the final Routh-Hurwitz condition \eqref{vhopfcondition} becomes an equality the polynomial \eqref{generalcharpoly5} has one pair of purely imaginary complex conjugate roots. Upon fixing values for $\omega_1$ and $\omega_2$ we may solve the Routh-Hurwitz conditions (with the final condition \eqref{vhopfcondition} an equality) polynomial by polynomial to find parameters in $(\varepsilon,a)$-parameter space where the system undergoes a Hopf bifurcation. 

For example, one set of conditions for a Hopf bifurcation of the trivial fixed point $P_0$ in the case where $\omega_1 = 1$ and $\omega_2 = -1$ is that $0.3 < \varepsilon < 0.437913$ and $a$ is either\footnote{depending on $\varepsilon$ it can be one or both} the third or fourth root\footnote{when the roots are ordered in increasing real part, with real roots listed before complex roots and complex conjugate pairs listed next to each other} of the polynomial:
\begin{align*}
&x^4 \left(3000000 \varepsilon^5-44050000 \varepsilon^4+37890000 \varepsilon^3-12964500 \varepsilon^2+2184300 \varepsilon-162000\right)\\
&+x^3 \left(6000000 \varepsilon^6-99900000 \varepsilon^5+111210000 \varepsilon^4-51363000 \varepsilon^3+12417300 \varepsilon^2\right.\\
&\left.-1634580 \varepsilon+97200\right)+x^2 \left(3000000 \varepsilon^7-66750000 \varepsilon^6+104805000 \varepsilon^5-123130000 \varepsilon^4\right.\\
&\left.+82932450 \varepsilon^3-27585675 \varepsilon^2+4710987 \varepsilon-338580\right)+x \left(-10900000 \varepsilon^7\right.\\
&\left.+32985000 \varepsilon^6-103331500 \varepsilon^5+102939450 \varepsilon^4-48149595 \varepsilon^3+12146787 \varepsilon^2\right.\\
&\left. -1649160 \varepsilon+97200\right)-900000 \varepsilon^6+10215000 \varepsilon^5-31817000 \varepsilon^4+29239350 \varepsilon^3\\
&-11819790 \varepsilon^2+2232900 \varepsilon-162000
\end{align*}
In particular we fix $\varepsilon = 0.31$ which gives two values for $a$ as the third and fourth root of the polynomial 
$$-127977700 - 1387299577 x + 1437004310065 x^2 - 152978250 x^3 - 6147350000 x^4$$
 so that $a \approx 0.00993214$ and $a \approx 15.2763$. So we have that the parameter sets $(\varepsilon,a) = (0.31,0.00993214)$ and $(\varepsilon,a) = (0.31,15.2763)$ result in  Hopf bifurcations of the trivial fixed point. We also note that the Routh-Hurwitz stability conditions at the two nontrivial fixed points are not satisfied for these parameters, so the other two fixed points do not bifurcate. 
 
 Alternatively for $\omega_1 = 1$ and $\omega_2 = -1$, solving the Routh-Hurwitz conditions for the nontrivial fixed points gives the range $1.60071<\varepsilon\leq1.82239$ where $a$, again, is a root of a polynomial whose coefficients depend on $\varepsilon$. For example, taking $\varepsilon= 1.65$, we obtain  $a \approx 0.0101494$ and $a\approx 4.20511$ as Hopf bifurcation points.


\subsection{Cyclically Coupled and Delayed Sprott System}

Next consider the Sprott system with cyclic coupling which is is given by
\begin{align}\label{undelaysprott}
\dot{x}_1 &= x_1 y_1 - \omega_1 z_1 + \varepsilon_1(x_2-x_1)\\
\dot{y}_1 &= x_1 - y_1\\
\dot{z}_1 &= \omega_1x_1 + \alpha z_1\\
\dot{x}_2 &= x_2 y_2 - \omega_2z_2\\
\dot{y}_2 &= x_2-y_2\\
\dot{z}_2 &= \omega_2 x_2 + \alpha z_2 + \varepsilon_2(z_1-z_2)
\end{align}
where $\alpha = 3/10$, and we note that each individual oscillator is chaotic in isolation for this value of $\alpha$.

Introducing a weak distributed time delay in the last equation, and considering the case where $\omega_1 = -\omega_2 = \omega$ and $\varepsilon_1 = \varepsilon_2 = \varepsilon$:
\begin{align}
\label{delaysprott1}
\dot{x}_1 &= x_1 y_1 - \omega z_1 + \varepsilon(x_2-x_1)\\
\dot{y}_1 &= x_1 - y_1\\
\dot{z}_1 &= \omega x_1 + \alpha z_1\\
\dot{x}_2 &= x_2 y_2 + \omega z_2\\
\dot{y}_2 &= x_2-y_2\\
\dot{z}_2 &= -\omega x_2 + \alpha z_2 + \varepsilon\left(\int_{-\infty}^t az_1(\tau)e^{-a(t-\tau)}d\tau-z_2\right)
\end{align}
where $\alpha = 3/10$, and by defining $w(t)$ as:
\begin{equation}
w(t) = \int_{-\infty}^t z_1(\tau)ae^{-a(t-\tau)}d\tau\notag
\end{equation}
we can reduce the system \eqref{delaysprott1} to the system of ordinary differential equations:
\begin{align}
\label{delaysprott}
\dot{x}_1 &= x_1 y_1 - \omega z_1 + \varepsilon(x_2-x_1)\\
\dot{y}_1 &= x_1 - y_1\\
\dot{z}_1 &= \omega x_1 + \alpha z_1\\
\dot{x}_2 &= x_2 y_2 + \omega z_2\\
\dot{y}_2 &= x_2-y_2\\
\dot{z}_2 &= -\omega x_2 + \alpha z_2 + \varepsilon\left(w-z_2\right)\\
\dot{w} &= a(z_1-w)
\end{align}

The fixed points of the delayed system are the trivial fixed point: \begin{equation}
P_0 =  (0, 0, 0, 0, 0,0,0)
\end{equation}
and the nontrivial fixed point:
\begin{equation}
    P_1 = \left( x_1^*, x_1^*, -\frac{\omega_1x_1^*}{\alpha}, x_2^*, x_2^*, \frac{(x_2^*)^2}{\omega_2}, -\frac{\omega_1x_1^*}{\alpha} \right)
\end{equation}
where
\begin{equation}\label{d-nontrivx2}
    x_2^* = \frac{1}{\varepsilon}\left(-\frac{\omega^2 x_1^*}{\alpha}+\varepsilon x_1^*-(x_1^*)^2\right)
\end{equation}
and $x_1^*$ is the real root of the cubic equations given by:
\begin{equation}\label{d-nontrivx1cond}
    \frac{\omega}{\varepsilon}\left( \varepsilon - \frac{\omega^2}{\alpha} - x_1^*\right)+  \frac{(\alpha-\varepsilon)x_1^*}{\omega\varepsilon_1^2}\left(\varepsilon + \frac{\omega^2}{\alpha} - x_1^*\right)^2+ \frac{\omega\varepsilon}{\alpha} = 0
\end{equation}

The eigenvalues of the Jacobian matrix of \eqref{delaysprott} at the trivial fixed point $P_0$ satisfy the characteristic equation
\begin{equation} 
\label{efunc2}
\lambda^7 + b_1\lambda^6 + b_2 \lambda^5 +b_3\lambda^4 + b_4\lambda^3 + b_5\lambda^2 + b_6\lambda + b_7 =0
\end{equation}
where the coefficients $b_i, i =1, 7$ are given in Appendix B.

Similarly, the eigenvalues of the Jacobian matrix at the fixed point $P_1$ satisfy the characteristic equation:
\begin{equation} 
\label{efunc2n}
\lambda^7 + b_1\lambda^6 + b_2 \lambda^5 +b_3\lambda^4 + b_4\lambda^3 + b_5\lambda^2 + b_6\lambda + b_7 =0
\end{equation}
where the coefficients $b_i, i = 1,7$ are also given in Appendix B.

For the fixed point $P_i$, $i=0,1$, to be  stable within the linearized analysis, all the eigenvalues must have negative real parts. From the Routh-Hurwitz criterion, the necessary and sufficient conditions for a seventh degree polynomial equation of the form:

\begin{equation}\label{generalcharpoly}
\lambda^7 + b_1\lambda^6 + b_2 \lambda^5 +b_3\lambda^4 + b_4\lambda^3 + b_5\lambda^2 + b_6\lambda + b_7 =0
\end{equation}

\noindent
to have $\hbox{Re} (\lambda_{1,2,3,4,5}) <0$ are:
\begin{align}
\label{srh1}
b_1 &>0 \\ 
\label{srh7}
b_7 &>0 \\ 
\label{srh2}
b_1 b_2 - b_3 &>0 \\ 
\label{srh3}
b_1 (b_2 b_3 + b_5) -b_3^2 - b_1^2 b_4 &>0 \\
\label{srh4}
-b_3^2 b_4-b_5^2+b_1^2 (-b_4^2+b_2 b_6)+b_3 (b_2 b_5+b_7)\quad&\notag\\ 
-b_1 (b_2^2 b_5-2 b_4 b_5+b_3 b_6+b_2 (-b_3 b_4+b_7)) &>0 \\
\label{srh5}
b_1^3 \left(-b_6^2\right)+b_1^2 \left(b_4 (b_3 b_6-b_2 b_7)+2 b_6 (b_2 b_5+b_7)-b_4^2 b_5\right)\quad&\notag\\
+b_1 \left(b_2^2 \left(b_3 b_7-b_5^2\right)-b_2 \left(b_3^2 b_6-b_3 b_4 b_5+b_5 b_7\right)-3 b_3 b_5 b_6+2 b_4 b_5^2-b_7^2\right)\quad&\notag\\
-b_3^2 (b_2 b_7+b_4 b_5)+b_3 b_5 (b_2 b_5+2 b_7)+b_3^3 b_6-b_5^3 &>0 \\ 
\label{shopfcondition}
-b_1^3 b_6^3+b_1^2 \left(b_4 b_6 (b_3 b_6-3 b_2 b_7)+b_6^2 (2 b_2 b_5+3 b_7)+b_4^3 b_7-b_4^2 b_5 b_6\right)\quad&\notag\\
-b_1 \left(b_2^3 b_7^2+b_2^2 \left(-2 b_3 b_6 b_7-b_4 b_5 b_7+b_5^2 b_6\right)+b_2 \left(b_3^2 b_6^2+b_3 b_4 (b_4 b_7-b_5 b_6)\right.\right.\quad&\notag\\
\left.\left.+b_7 (b_5 b_6-3 b_4 b_7)\right)-b_4 b_6 \left(b_3 b_7+2 b_5^2\right)+3 b_6 \left(b_3 b_5 b_6+b_7^2\right)+2 b_4^2 b_5 b_7\right)\quad&\notag\\
+b_3 \left(b_2^2 b_7^2+b_2 b_5 (b_5 b_6-b_4 b_7)+b_7 (3 b_5 b_6-2 b_4 b_7)\right)+b_3^2 \left(-2 b_2 b_6 b_7+b_4^2 b_7\right.\quad&\notag\\
\left.-b_4 b_5 b_6\right)-b_2 b_5 b_7^2+b_3^3 b_6^2+b_4 b_5^2 b_7-b_5^3 b_6+b_7^3&>0
\end{align}

It is straightforward to check that, for $\alpha = 3/10$, the fixed point $P_0$ undergoes a transcritical bifurcation, colliding and exchanging stability with $P_1$, when
\begin{equation} \label{transcritical}
10 \varepsilon^2 - 3 \varepsilon + 10 \omega^2  = 0
\end{equation}
For $ = 3/10$ and $\varepsilon_1 = \varepsilon_2 = \varepsilon$, this transcritical bifurcation curve is plotted below in Figure \ref{fig:transcritical}

\begin{figure}
\centering
\centerline{\includegraphics[width=0.5\textwidth]{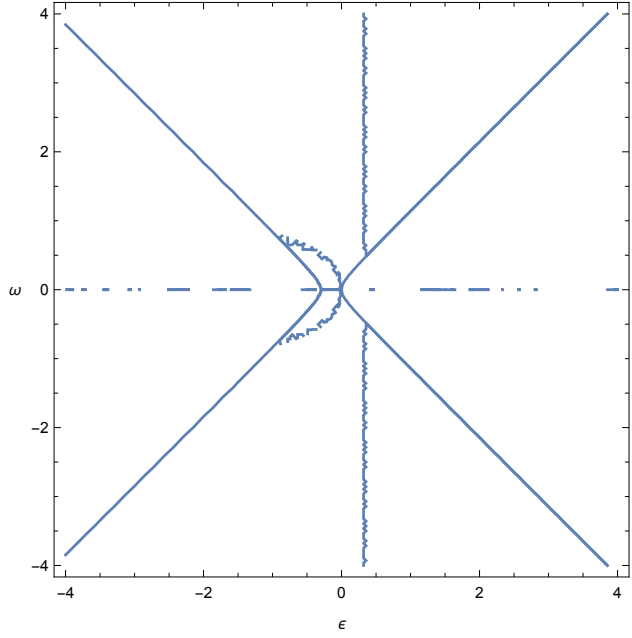}}
\caption{\label{fig:transcritical} Transcritical bifurcation curves where fixed points $P_0$ and $P_1$ of \eqref{delaysprott} collide and exchange stability for $\alpha = 3/10$. The relevant portions are to the {\it right} of the intersections of the jagged curves with both the left and right rotated V-shaped curves .}
\end{figure}

When the final condition \eqref{shopfcondition} becomes an equality, the polynomial \eqref{generalcharpoly} has one pair of purely imaginary complex conjugate roots. Upon fixing values for $\omega$ we may numerically solve the Routh-Hurwitz conditions with the final condition \eqref{shopfcondition} taken to be equality (along with equation \eqref{d-nontrivx1cond} for $x_1^*$ in the nontrivial case) to find parameters in $(\varepsilon,a)$-parameter space where the system undergoes a Hopf bifurcation. 

For example, one set of conditions for a Hopf bifurcation of the trivial fixed point $P_0$ we obtain in the case where $\omega=5$ is that $0.3 < \varepsilon < 0.457474$ and $a$ is either the third or fourth root\footnote{when the roots are ordered in increasing real part, with real roots listed before complex roots and complex conjugate pairs listed next to each other} of the polynomial:
\begin{align*}
&x^4 \left(3000000 \varepsilon^5-1004050000 \varepsilon^4+901890000 \varepsilon^3-315364500 \varepsilon^2+54024300 \varepsilon-4050000\right)\\
&\quad+x^3 \left(6000000 \varepsilon^6-2259900000 \varepsilon^5+2631210000 \varepsilon^4-1239363000 \varepsilon^3+304017300 \varepsilon^2\right.\\
&\quad\left.-40514580 \varepsilon+2430000\right)+x^2 \left(3000000 \varepsilon^7-1506750000 \varepsilon^6+2480805000 \varepsilon^5\right.\\
&\quad\left.-36019930000 \varepsilon^4+38060772450 \varepsilon^3-14728329675 \varepsilon^2+2709722187 \varepsilon-202864500\right)\\
&\quad+x \left(-250900000 \varepsilon^7+788985000 \varepsilon^6-44560931500 \varepsilon^5+54766779450 \varepsilon^4\right.\\
&\quad\left.-27662777595 \varepsilon^3+7269617187 \varepsilon^2-1013229000 \varepsilon+60750000\right)-22500000 \varepsilon^6\\
&\quad+5655375000 \varepsilon^5-319545425000 \varepsilon^4+378208983750 \varepsilon^3-169382994750 \varepsilon^2\\
&\quad+33795562500 \varepsilon-2531250000
\end{align*}
In particular, we can fix $\varepsilon = 0.4$ which gives $a$ as the third and fourth root of the polynomial $-31473200 - (16371964/5)x + 57507566 x^2 - 210120 x^3 - 850600 x^4$, or $a \approx 0.773509$ and $a \approx8.03562$. So we have that the parameter sets $(\varepsilon,a) = (0.4,0.773509)$ and $(\varepsilon,a) = (0.4,8.03562)$ result in Hopf bifurcations of the trivial fixed point. Here we note that the Routh-Hurwitz stability conditions at the nontrivial fixed point are satisfied for these values of $\omega$ and $\varepsilon$ for any choice of $a$, so $P_1$ does not bifurcate at these parameters as we vary $a$ around the Hopf bifurcation point of $P_0$. 
 
 Alternatively for $\omega = 1$ and $\varepsilon=1.41$, we can numerically solve the Routh-Hurwitz conditions and condition \eqref{d-nontrivx1cond} on $x_1^*$ for the nontrivial fixed point to obtain that $x_1^* \approx 0.722446$, so that the nontrivial fixed point is 
 $$P_1\approx(0.722446, 0.722446, -2.40815, -1.35563, -1.35563, -1.83772, -2.40815)$$
 and $a \approx 0.241658$ and $a\approx 23.8302$ as bifurcation points. Here, we again note that for these values of $\omega$ and $\varepsilon$ the the Routh-Hurwitz conditions at the trivial fixed point are not satisfied for any choice of $a$ and so it does not undergo an Hopf bifurcation as we vary $a$ around the bifurcation points of the nontrivial fixed point.

\section{Numerical Results and Discussion}

\subsection{Delayed Van der Pol System}
Let us now turn to numerical results for the Van Der Pol System. Here we will consider two sets of parameters: one for $\omega_1 = 1$ and $\omega_2 = -1$, and the other for $\omega_1 = \omega_2 = 1$, corresponding to counter- and co-rotating oscillators respectively. In general the 'parameter mismatch' $\Delta = \omega_2/\omega_1$ allows for symmetry breaking of the system via a pitchfork bifurcation, as already discussed earlier, and plotted in Figure 1.

\subsubsection{Parameter Set 1 ($\omega_1 =-\omega_2=1$):  Trivial Fixed Point}

Here we will consider the case where $\omega_1 = 1$, $\omega_2 = -1$, $\varepsilon = 31/100$ corresponding to the trivial fixed point being stable, i.e., prior to the symmetry-breaking pitchfork bifurcation
. First we note that for this set of parameters each Van der Pol system is in oscillation in isolation (that is uncoupled and without delay), while the coupled system (without delay) is in a state of amplitude death (that is the trivial fixed point is stable). For this set of parameters, the Routh Hurwitz conditions at the trivial fixed point show that the trivial fixed point Hopf bifurcates at $a\approx 0.00993214$ and $a\approx 15.2763$. By contrast, the nontrivial fixed points do not bifurcate as we vary $a$ for this case.

\begin{figure}[H]
\centering
\centerline{\includegraphics[width=0.6\textwidth]{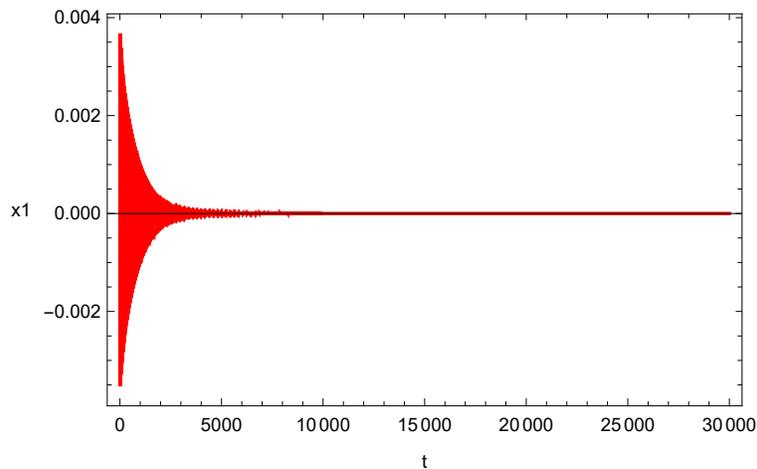}}
\caption{\label{fig:vdp-a20-x1}Amplitude Death in $x_1$ for $a = 20$.}
\end{figure}

\begin{figure}
\centering
\centerline{\includegraphics[width=0.5\textwidth]{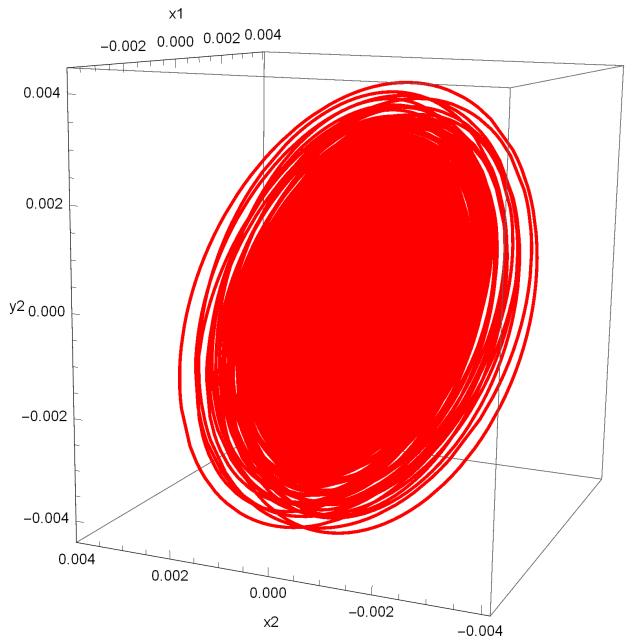}}
\caption{\label{fig:vdp-a20-x1x2y2} The solution for $a=20$ spiraling in towards the stable origin from the initial conditions.}
\end{figure}

Figure \ref{fig:vdp-a20-x1} shows the solutions for $x_1$ for $a = 20$ above the first Hopf bifurcation value $a\approx 15.2763$. Here, the origin is stable and we have amplitude death above the first bifurcation point. Figure \ref{fig:vdp-a20-x1x2y2} shows the solution in $(x_1,x_2,y_2)$ phase space and the approach from the initial conditions as the solution spirals towards the origin. 

\begin{figure}
\centering
\centerline{\includegraphics[width=0.7\textwidth]{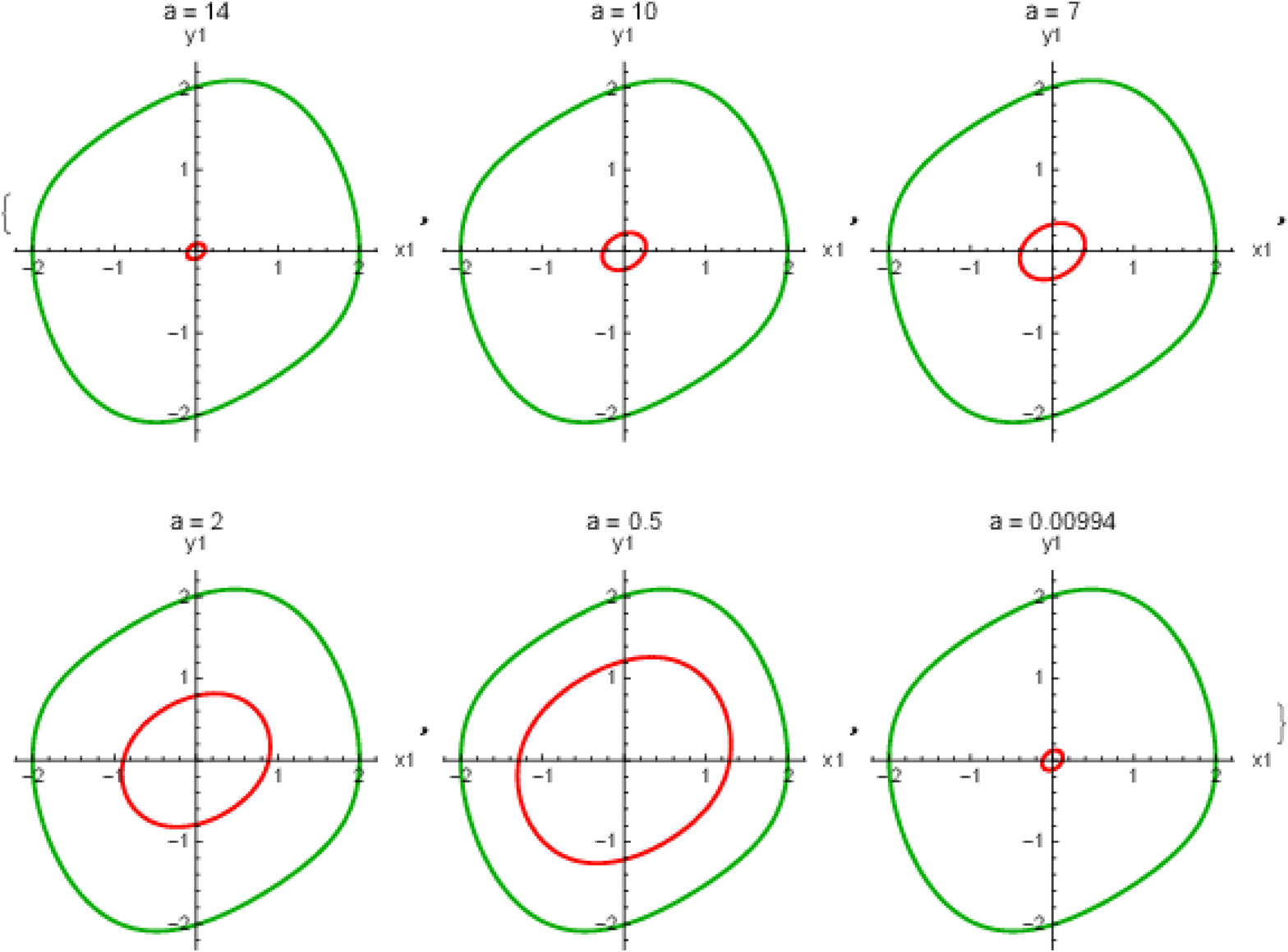}}
\caption{\label{fig:vdp-periodic-compare} The limit cycle of an isolated undelayed Van der Pol oscillator in green and the limit cycle of the delayed system in red for various values of $a$ between the two bifurcation points.  }
\end{figure}

In figure \ref{fig:vdp-periodic-compare} we have plotted the limit cycle of the isolated (undelayed) Van der Pol oscillator in green and the solutions in $(x_1,y_1)$ phase space of the delayed, coupled system (in red) for various values of $a$ between the two bifurcation points of our system. We observe, as expected, that on this side of the Hopf bifurcation point we have periodic behavior. Also, just below the first bifurcation point $a\approx 15.2763$ the limit cycle is very small and close to the origin and as we decrease the delay parameter $a$ the limit cycle grows in size. Then as we start to approach the second bifurcation point $a\approx 0.00993214$, we see that immediately above $a = 0.00994$ the delayed limit cycle has begun to shrink toward the origin again. 

Next  Figures \ref{fig:vdp-a0-005-x1} shows the delayed solution for $a=0.005$ below the second bifurcation value $a\approx 0.00993214$. Here we see that below the second Hopf bifurcation point the delayed system experiences amplitude death as the origin regains stability. In Figure \ref{fig:vdp-a0.005-x1x2y1} we see the solution in $(x_1,x_2,y_1)$ parameter space approaching the now stable origin from the initial conditions. 

\begin{figure}
\centering
\centerline{\includegraphics[width=0.6\textwidth]{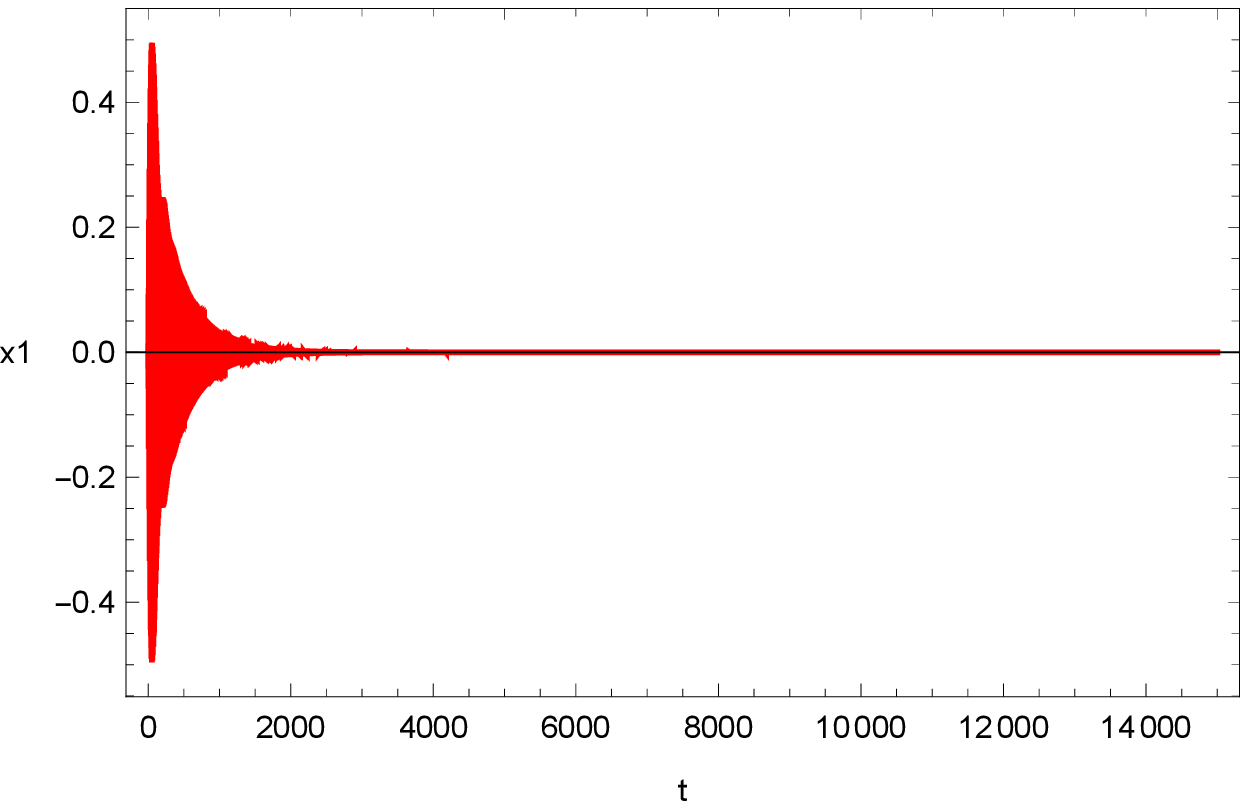}}
\caption{\label{fig:vdp-a0-005-x1} Amplitude Death in $x_1$ for $a = 0.005$. }
\end{figure}

\begin{figure}
\centering
\centerline{\includegraphics[width=0.5\textwidth]{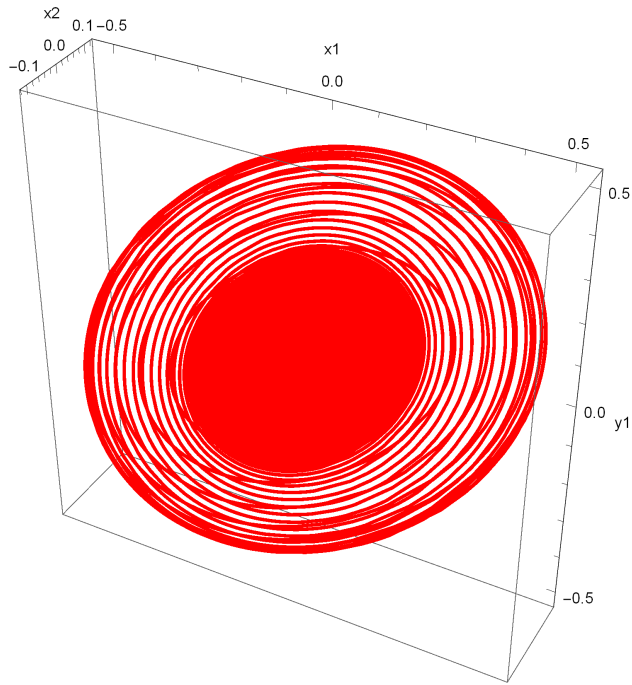}}
\caption{\label{fig:vdp-a0.005-x1x2y1} The solution in $(x_1,x_2,y_1)$ parameter space for $a = 0.005$ spiraling towards the stable origin from the initial conditions. }
\end{figure}

\subsubsection{Parameter Set 1 ($\omega_1 = -\omega_2 =1$): Nontrivial Fixed Points}

Here we will consider the case where $\omega_1 = 1$, $\omega_2 = -1$, $\varepsilon = 1.65$. As we can see from \eqref{pitchfork}, for $b = 3/10$ the trivial fixed point undergoes a pitchfork bifurcation at $\varepsilon = 0.862$, and so we are now past that bifurcation where the stable nontrivial fixed points were born. First we note that for this set of parameters each Van der Pol system is in oscillation in isolation (that is uncoupled and without delay), while the coupled system (without delay) is in a state of amplitude death (that is the trivial fixed point is stable). For this set of parameters the Routh Hurwitz conditions at the nontrivial fixed point show that it bifurcates at $a\approx 0.0101494$ and $a\approx 4.20511$. The Routh Hurwitz conditions at the trivial fixed point show that, for our choice of $(\omega_1,\omega_2,\varepsilon)$, $P_0$ does not bifurcate as we vary $a$. For these parameters the two nontrivial fixed points are given by:
\begin{align}
    P_+ &\approx (2.11655, -2.02747, 3.34532, 0, -2.02747)\\ 
    P_- &\approx (-2.11655, 2.02747, -3.34532, 0, 2.02747)
\end{align}

\begin{figure}
\centering
\centerline{\includegraphics[width=0.5\textwidth,height=0.5\textwidth]{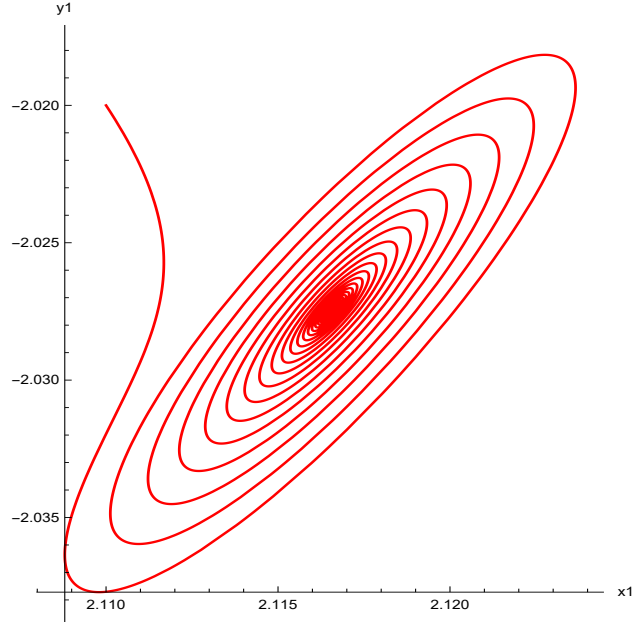}}
\caption{\label{fig:vdpn1-a9-phase1}As the coupled system approaches $P_+$, the first oscillator $(x_1,y_1)$ approaches the steady state $(2.11655, -2.02747)$.}
\end{figure}

\begin{figure}
\centering
\centerline{\includegraphics[width=0.5\textwidth,height=0.5\textwidth]{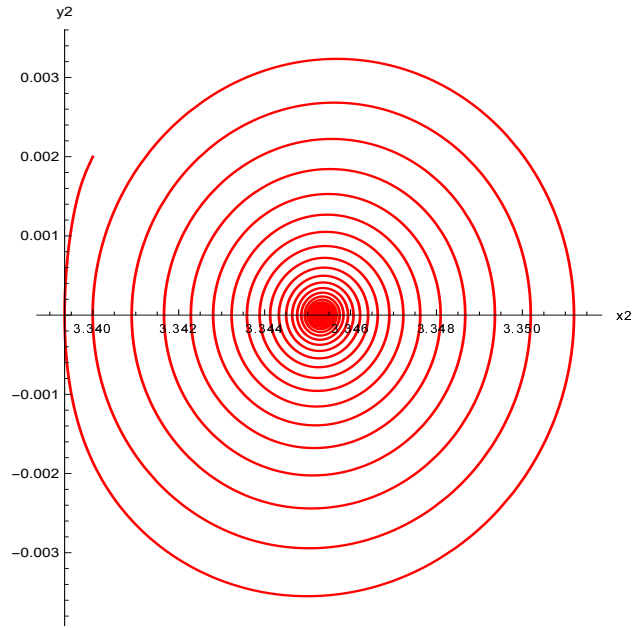}}
\caption{\label{fig:vdpn1-a9-phase2}As the coupled system approaches $P_+$, the second oscillator $(x_2,y_2)$ approaches the steady state $(3.34532, 0)$.}
\end{figure}

\begin{figure}
\centering
\centerline{\includegraphics[width=0.5\textwidth,height=0.5\textwidth]{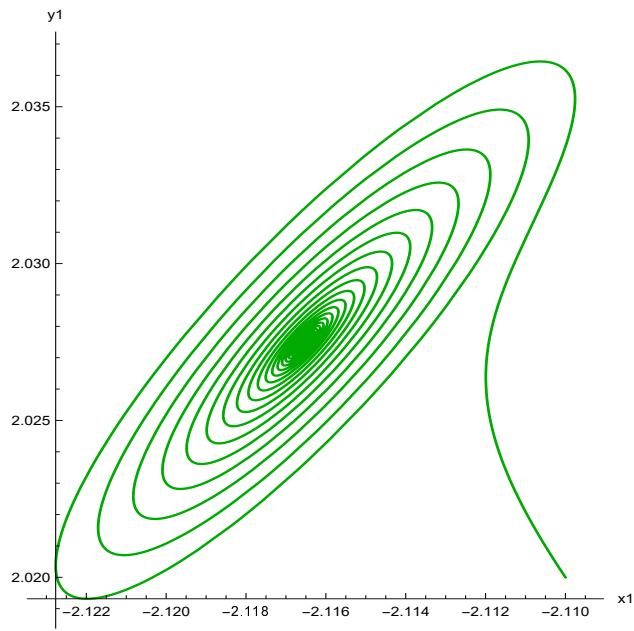}}
\caption{\label{fig:vdpn2-a9-phase1}As the coupled system approaches $P_-$, the first oscillator $(x_1,y_1)$ approaches the steady state $(-2.11655, 2.02747)$.}
\end{figure}

\begin{figure}
\centering
\centerline{\includegraphics[width=0.5\textwidth,height=0.5\textwidth]{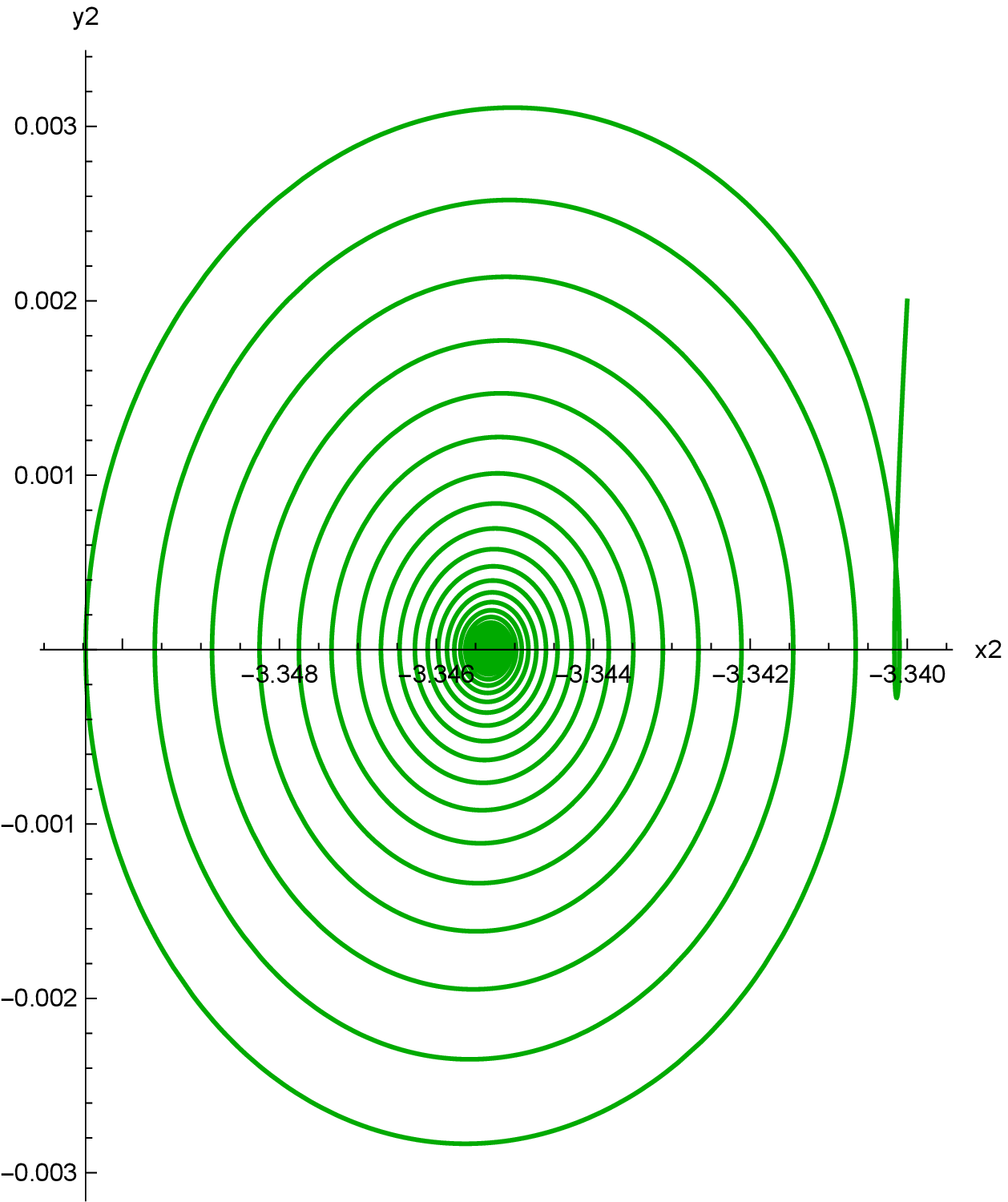}}
\caption{\label{fig:vdpn2-a9-phase2}As the coupled system approaches $P_-$, the second oscillator $(x_2,y_2)$ approaches the steady state $(-3.34532, 0)$.}
\end{figure}

Figure \ref{fig:vdpn1-a9-phase1}  shows the solution in phase space for the first oscillator $(x_1,y_1)$ and Figure \ref{fig:vdpn1-a9-phase2} shows the solution for the second oscillator in $(x_2,y_2)$ phase space with initial condition near $P_+$ for $a = 9$ above the first Hopf bifurcation value $a\approx 4.20511$. Similarly, Figure \ref{fig:vdpn2-a9-phase1}  shows the solution in phase plane for the first oscillator $(x_1,y_1)$ and Figure \ref{fig:vdpn2-a9-phase2} shows the solution for the second oscillator in $(x_2,y_2)$ phase space with initial condition near $P_-$ for $a = 9$. Here, both nontrivial fixed points are stable and we see that we have oscillation death above the first bifurcation point (that is two oscillators $(x_1,y_1)$ and $(x_2,y_2)$ settling to two distinct steady states). 

\begin{figure}
\centering
\centerline{\includegraphics[width=0.6\textwidth]{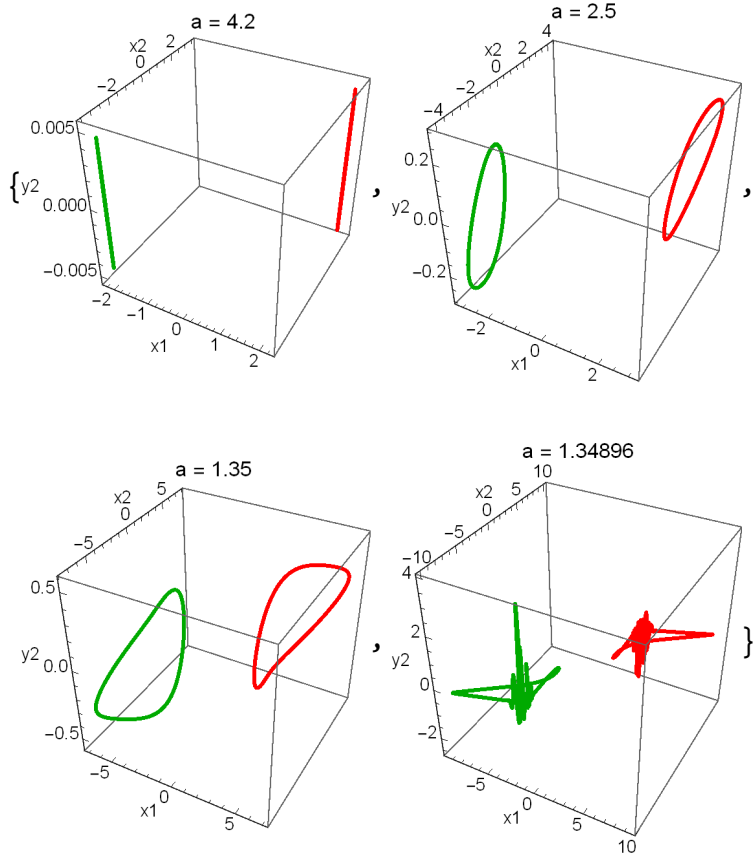}}
\caption{\label{fig:vdpn-3dcompare} The first three plots show the limit cycles for the delayed system for initial conditions near $P_+$ in red and initial conditions near $P_-$, for various values of $a$ between the two bifurcation points. The final plot for $a = 1.3486$ shows the solutions beginning to grow.}
\end{figure}

\begin{figure}
\centering
\centerline{\includegraphics[width=0.6\textwidth]{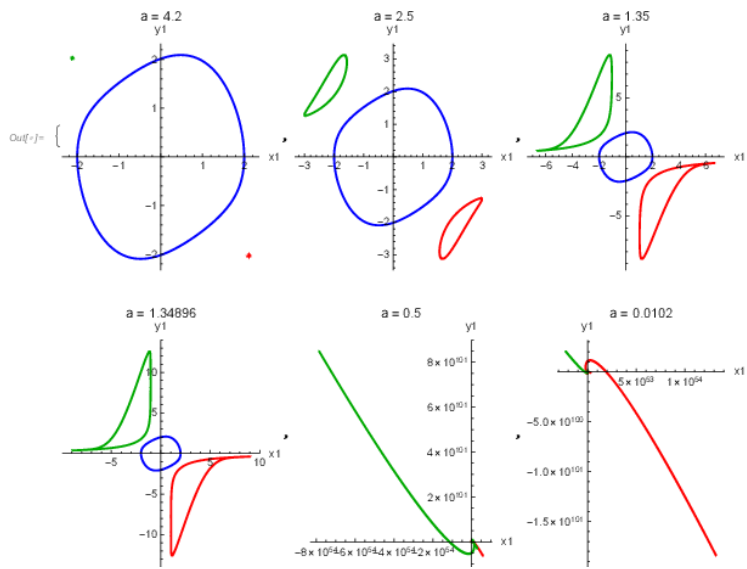}}
\caption{\label{fig:vdpn-2dcompare} The first four plots contain the limit cycles for the delayed system for initial conditions near $P_+$ in red and initial conditions near $P_-$, and the limit cycle of the undelayed and uncoupled system, for various values of $a$ between the two bifurcation points. The last two show that, as we further decrease $a$, we no longer have a stable limit cycle and the solutions fly off to infinity.}
\end{figure}

After the first Hopf bifurcation at $a\approx 4.20511$, both nontrivial fixed points become unstable. In figure \ref{fig:vdpn-3dcompare} we have plotted the limit cycles for the delayed system for initial conditions near $P_+$ in red, and initial conditions near $P_-$ in green in $(x_1,x_2,y_2)$ phase space for various values of $a$ between the first and second bifurcation points. Here we see that the limit cycle is stable and expands in size as we decrease the delay parameter $a$ until $a\approx 1.34896$ where the solutions begin to grow in size. In figure \ref{fig:vdpn-2dcompare} we plot the limit cycles for initial conditions starting near $P_+$ in red, and initial conditions starting near $P_-$ in green for the delayed system, and the limit cycle for an isolated, undelayed system in blue in the first four graphs. We see that the delayed limit cycles start out very small around each nontrivial fixed point and, as we decrease the delay parameter $a$, the limit cycles grow in size and begin to stretch out. Then, as we move even closer to the second bifurcation point (past $a\approx 1.347$), we observe that the solutions no longer tend to a stable limit cycle and instead solutions extend off towards infinity in the cases for $a = 0.5, 0.0102$. 

\begin{figure}
\centering
\centerline{\includegraphics[width=0.5\textwidth,height=0.5\textwidth]{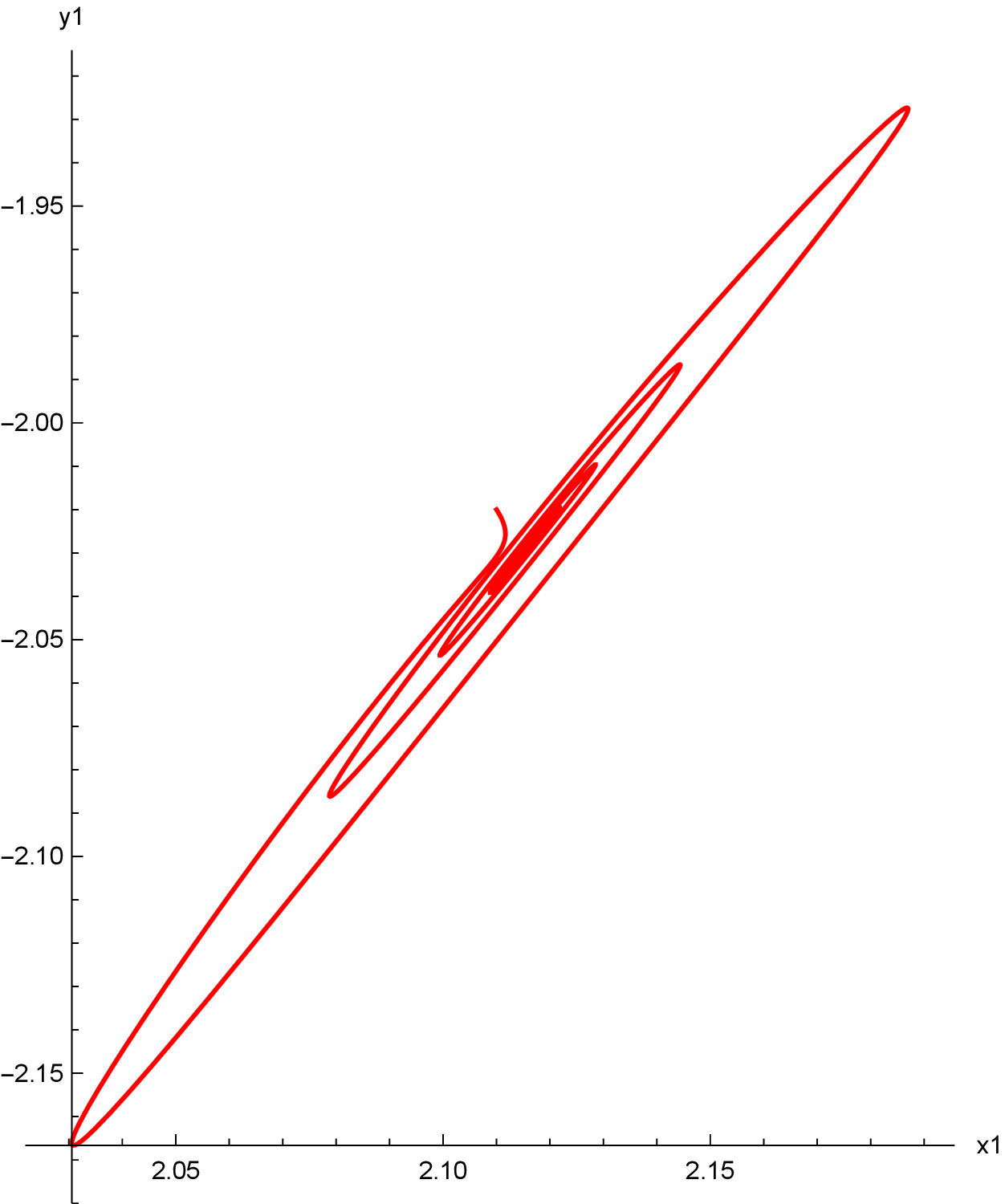}}
\caption{\label{fig:vdpn1-a0-005-phase1}As the coupled system approaches $P_+$, the first oscillator $(x_1,y_1)$ approaches the steady state $(2.11655, -2.02747)$.}
\end{figure}

\begin{figure}
\centering
\centerline{\includegraphics[width=0.5\textwidth,height=0.5\textwidth]{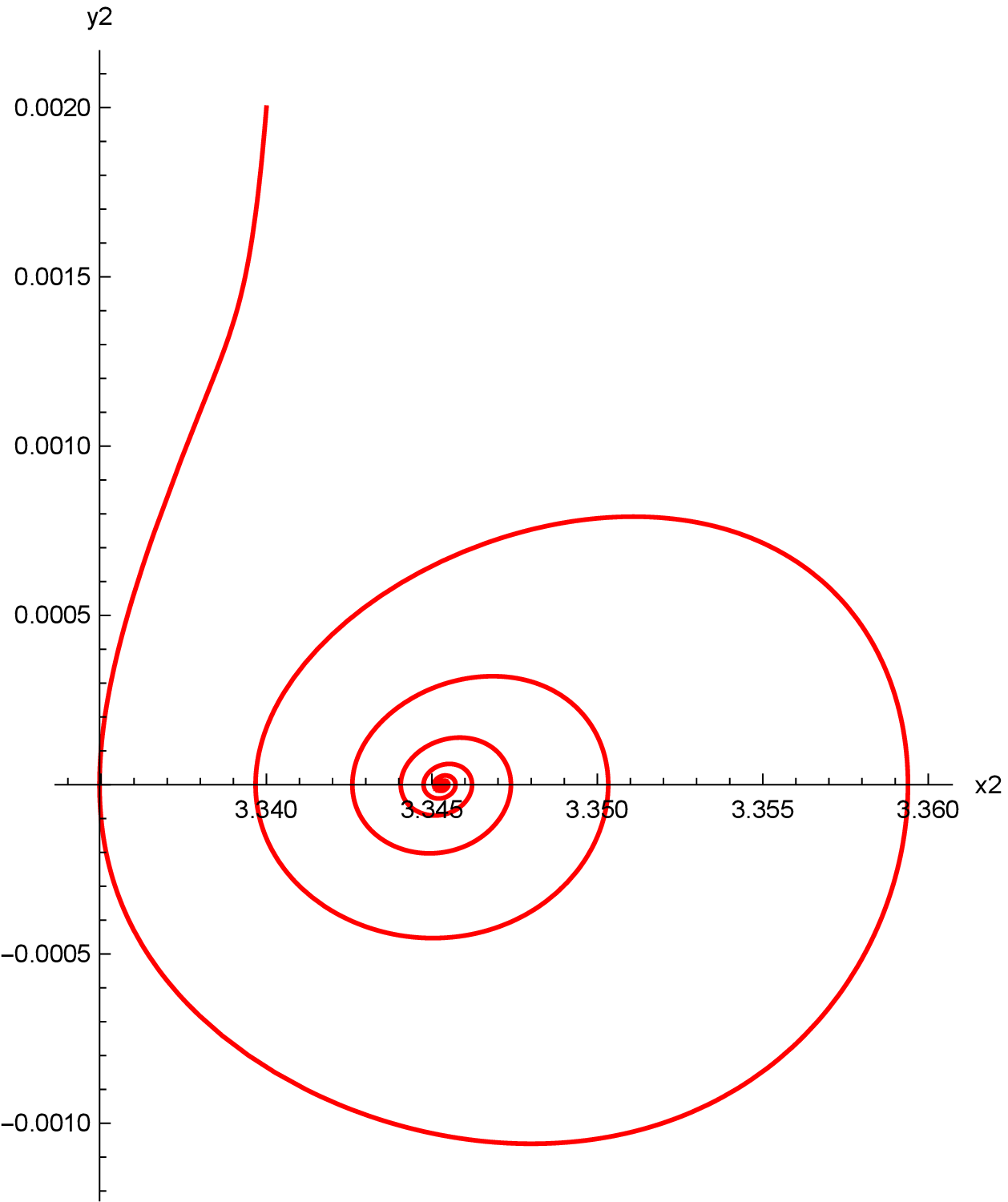}}
\caption{\label{fig:vdpn1-a0-005-phase2}As the coupled system approaches $P_+$, the second oscillator $(x_2,y_2)$ approaches the steady state $(3.34532, 0)$.}
\end{figure}

\begin{figure}
\centering
\centerline{\includegraphics[width=0.5\textwidth,height=0.5\textwidth]{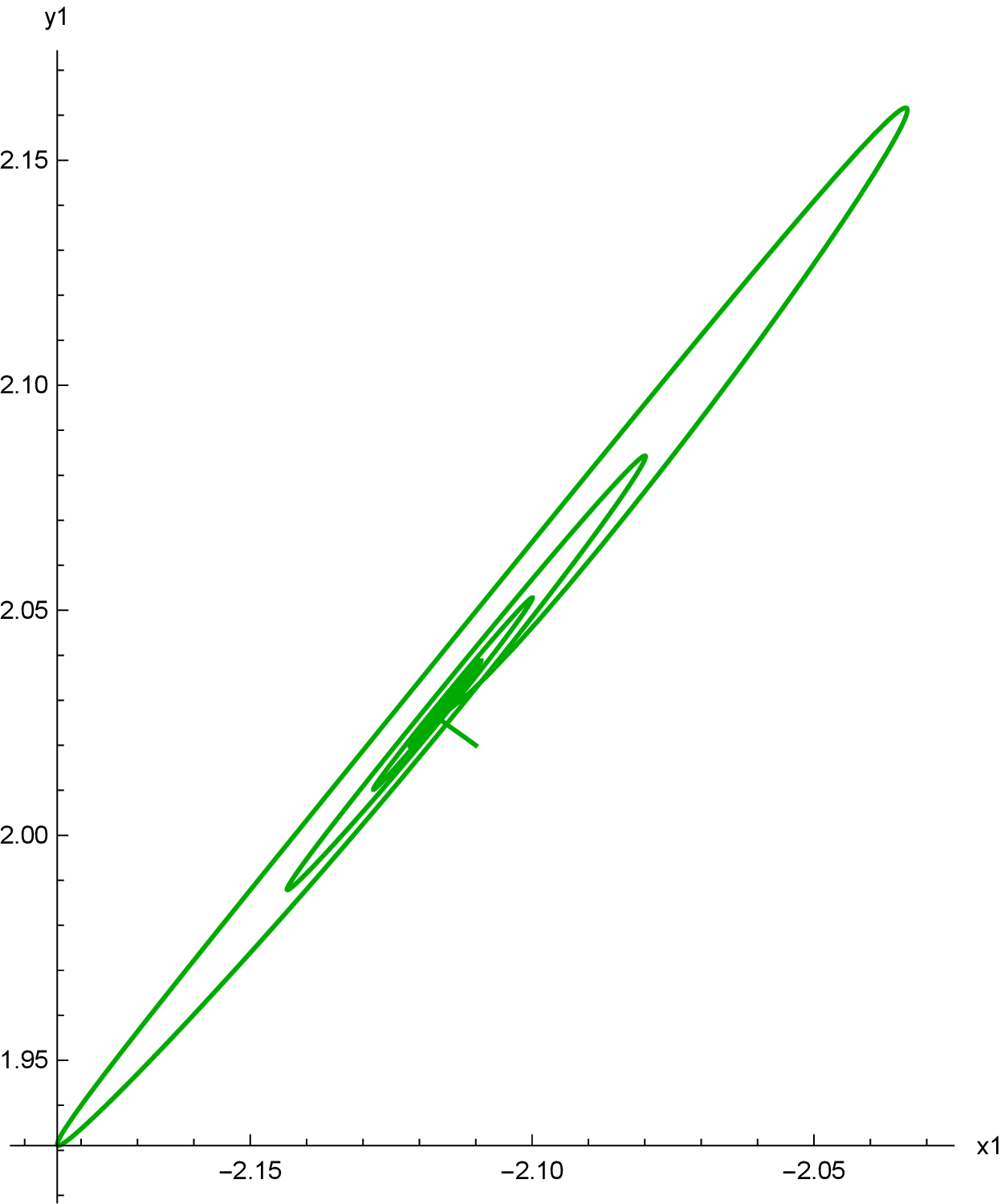}}
\caption{\label{fig:vdpn2-a0-005-phase1}As the coupled system approaches $P_-$, the first oscillator $(x_1,y_1)$ approaches the steady state $(-2.11655, 2.02747)$.}
\end{figure}

\begin{figure}
\centering
\centerline{\includegraphics[width=0.5\textwidth,height=0.5\textwidth]{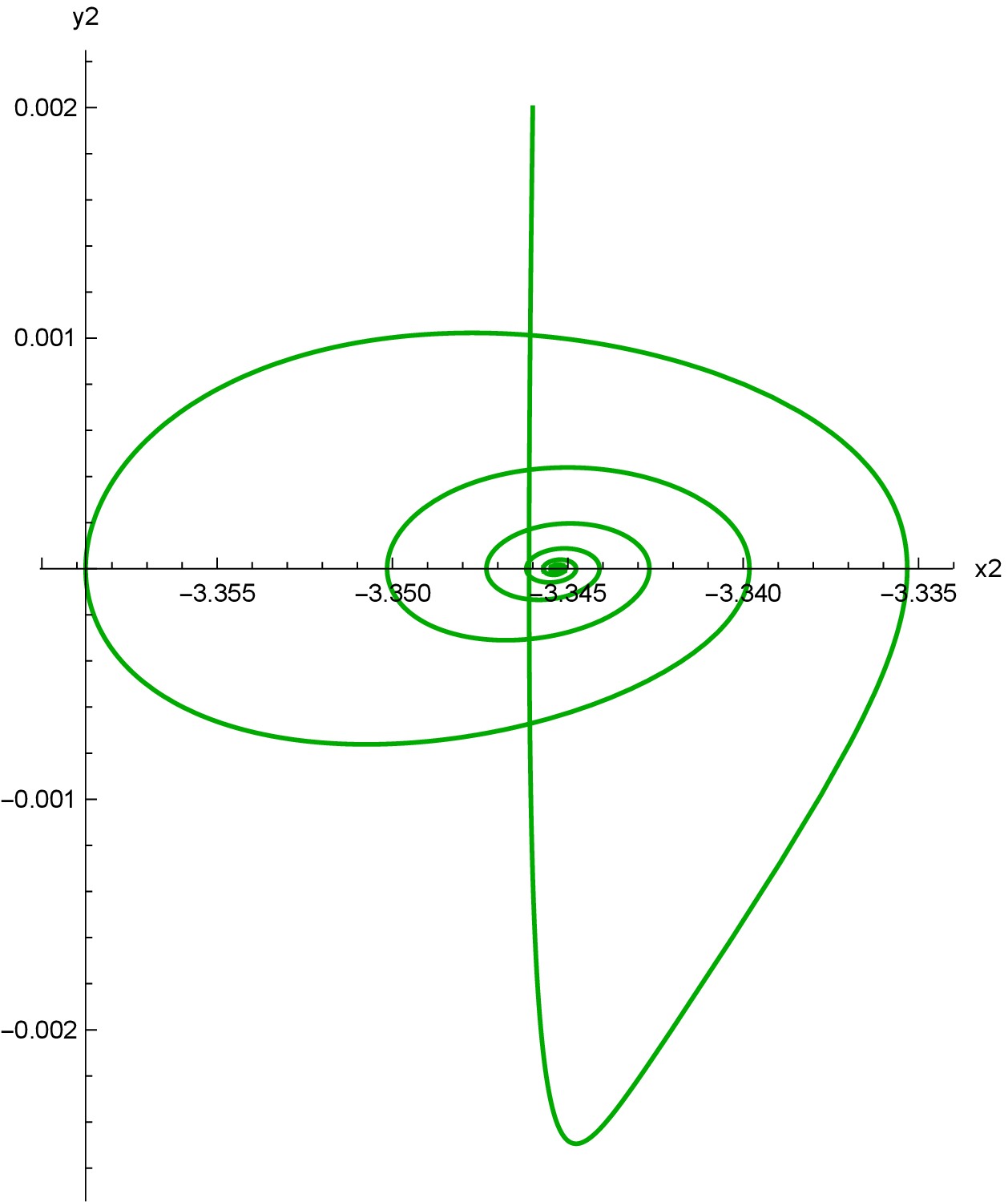}}
\caption{\label{fig:vdpn2-a0-005-phase2}As the coupled system approaches $P_-$, the second oscillator $(x_2,y_2)$ approaches the steady state $(-3.34532, 0)$.}
\end{figure}

Next we consider the delayed solution for $a=0.005$ below the second bifurcation value $a\approx  0.0101494$. Figure \ref{fig:vdpn1-a0-005-phase1}  shows the solution in phase plane for the first oscillator $(x_1,y_1)$ and Figure \ref{fig:vdpn1-a0-005-phase2} shows the solution for the second oscillator in $(x_2,y_2)$ phase space with initial condition near to $P_+$. Similarly, Figure \ref{fig:vdpn2-a0-005-phase1}  shows the solution in phase plane for the first oscillator $(x_1,y_1)$ and Figure \ref{fig:vdpn2-a0-005-phase2} shows the solution for the second oscillator in $(x_2,y_2)$ phase space with initial condition near to $P_-$. Here, both nontrivial fixed points have regained their stability, and we see that we have oscillation death below the second bifurcation point (that is two oscillators $(x_1,y_1)$ and $(x_2,y_2)$ settling to two distinct steady states).

\subsubsection{Parameter Set 2 ($\omega_1 = \omega_2 = 1$)}

Here we will consider the case where $\omega_1 = \omega_2 = 1$, $\varepsilon = 1$. First we note that for this set of parameters each Van der Pol system is in oscillation in isolation (that is uncoupled and without delay) and the coupled system (without delay) is in a state of oscillation as well. For this set of parameters the Routh Hurwitz conditions at the trivial fixed point give us that the trivial fixed point Hopf bifurcates at $a\approx 0.28607$, and the other two nontrivial fixed points do not exist as the pitchfork bifurcation boundary \eqref{pitchfork} has not been crossed.

\begin{figure}
\centering
\centerline{\includegraphics[width=0.6\textwidth]{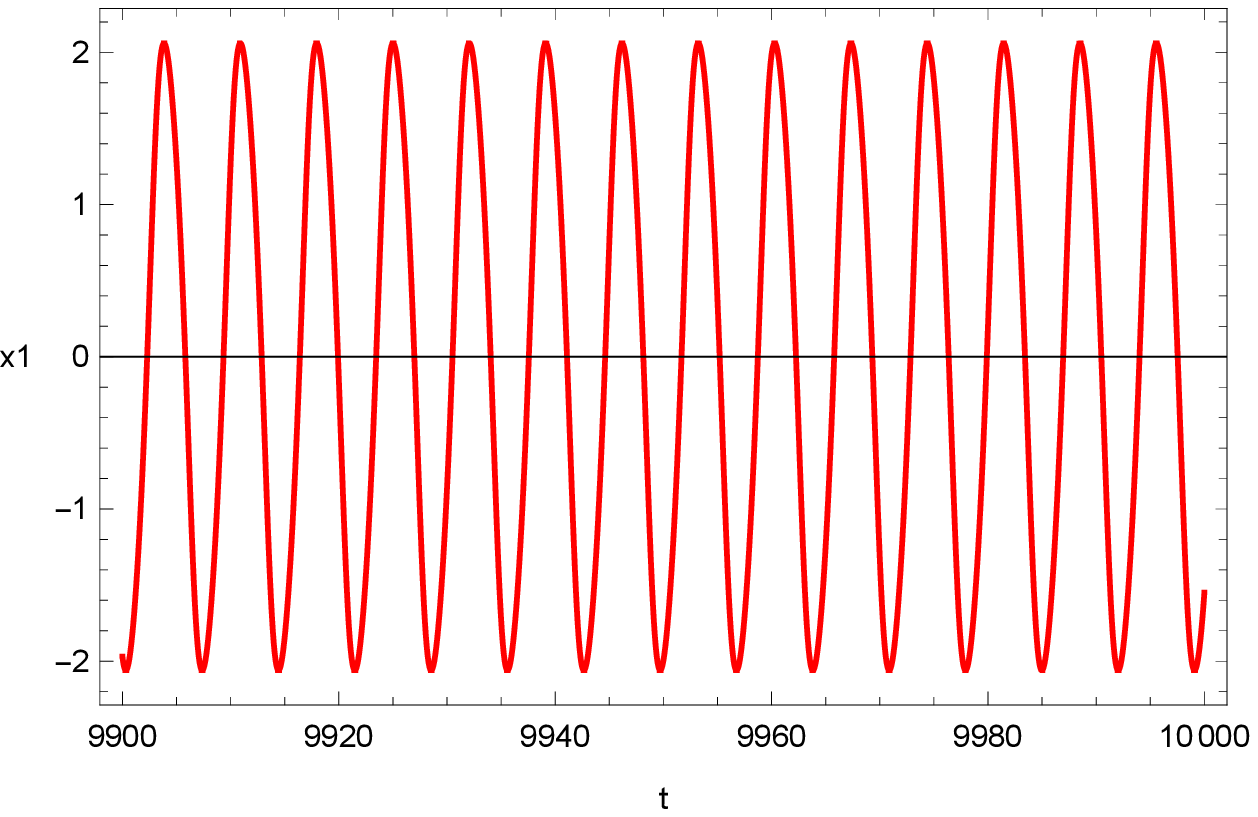}}
\caption{\label{fig:vdp2-a2-x1}Oscillations in $x_1$ for $a = 2$.}
\end{figure}

\begin{figure}
\centering
\centerline{\includegraphics[width=0.5\textwidth]{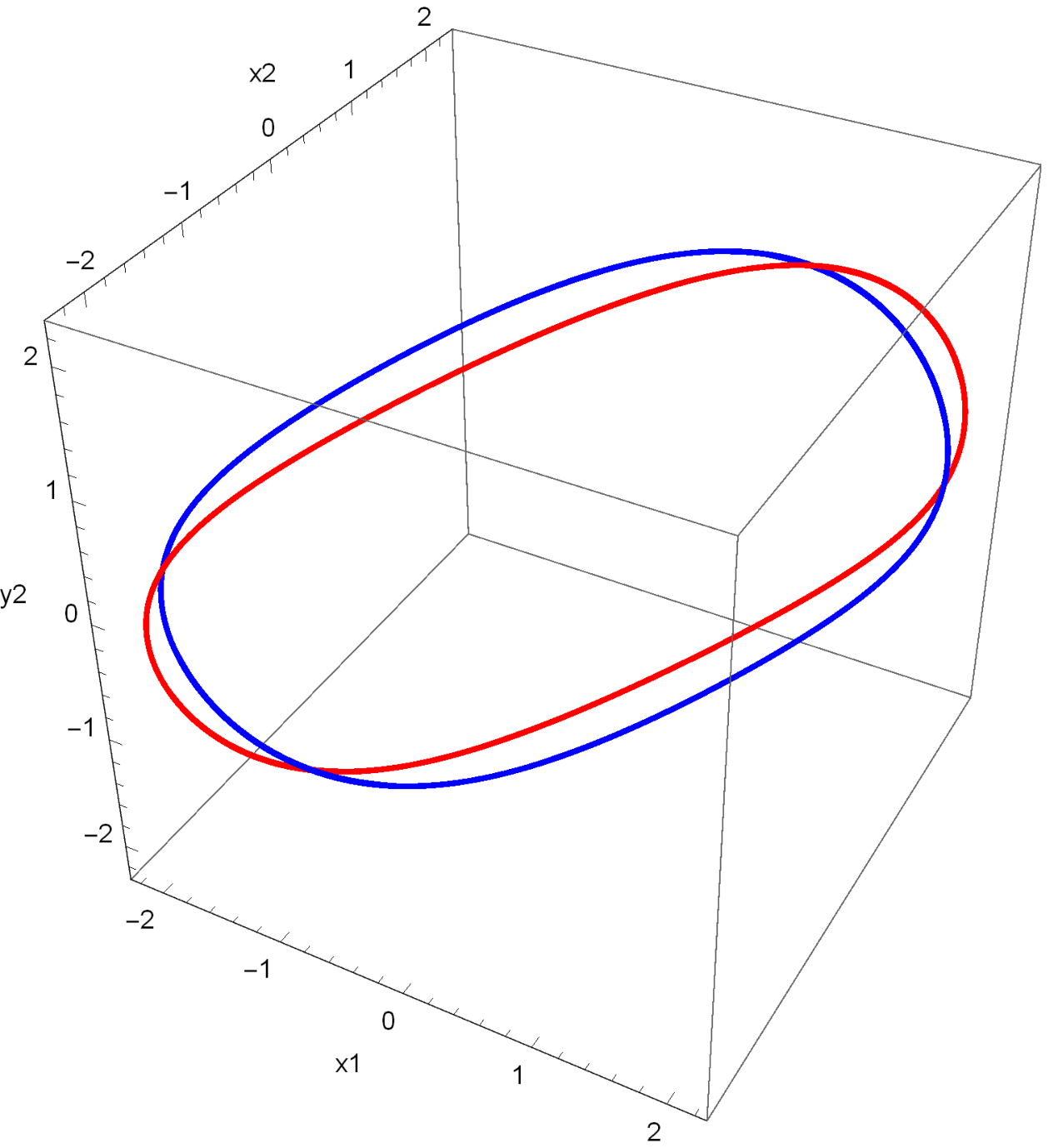}}
\caption{\label{fig:vdp2-a2-phase}Limit cycles for the coupled system without delay in blue, and for the coupled system with delay in red for $a = 2$ in $(x_1,x_2,y_2)$ phase space.}
\end{figure}

\begin{figure}
\centering
\centerline{\includegraphics[width=0.6\textwidth]{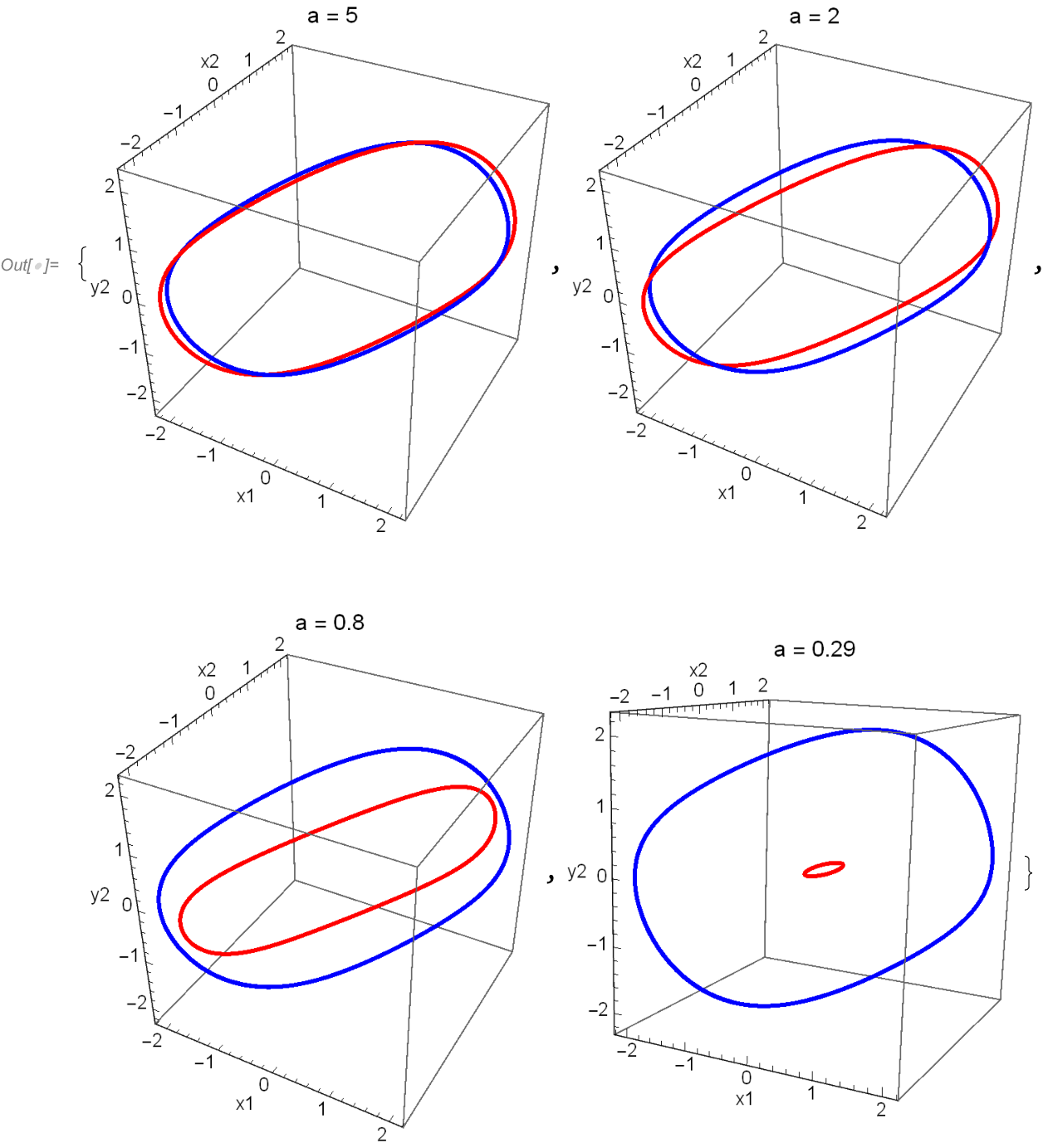}}
\caption{\label{fig:vdp2-compare} The limit cycles for the undelayed (in blue) and delayed (in red) systems for several values of the delay parameter $a$. We observe that the delayed limit cycle shrinks to the origin as we decrease $a$ towards the bifurcation point.}
\end{figure}

Figures \ref{fig:vdp2-a2-x1} shows the solution for $x_1$ for $a = 2$ above the Hopf bifurcation value $a\approx 0.28607$. Above the bifurcation point we have oscillatory behavior and Figure \ref{fig:vdp2-a2-phase} shows the limit cycles for the coupled system without delay in blue and the coupled system with delay in red in $(x_1,x_2,y_2)$ phase space. We see that the delay deforms and stretches the limit cycle. Figure \ref{fig:vdp2-compare} shows the undelayed (blue) and delayed (red) limit cycles for several values of the delay parameter $a$, from which we see that as we decrease $a$ towards the bifurcation point (that is strengthen the delay) the delayed limit cycle in red deforms from the undelayed one, becoming thinner and shrinking towards the origin. 

\begin{figure}
\centering
\centerline{\includegraphics[width=0.6\textwidth]{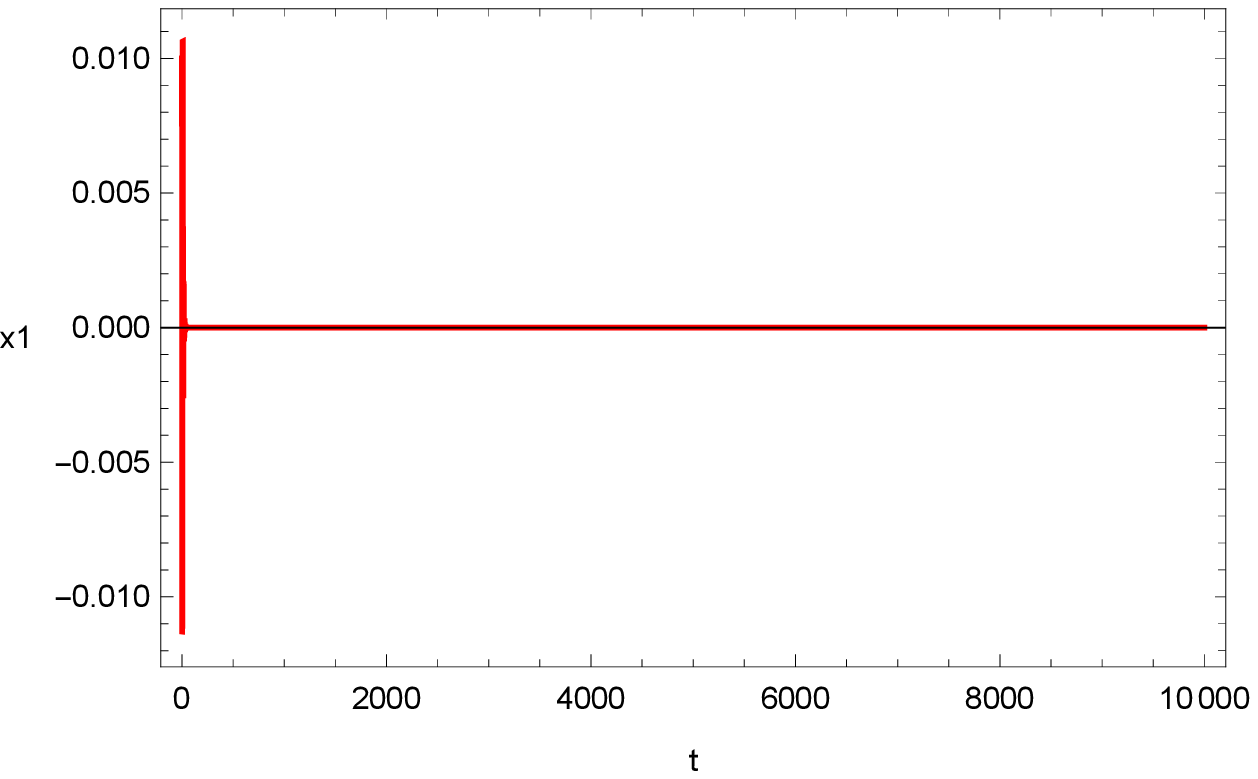}}
\caption{\label{fig:vdp2-a0-08-x1full} The solution for $x_1$ for the case $a = 0.08$.  }
\end{figure}

\begin{figure}
\centering
\centerline{\includegraphics[width=0.5\textwidth]{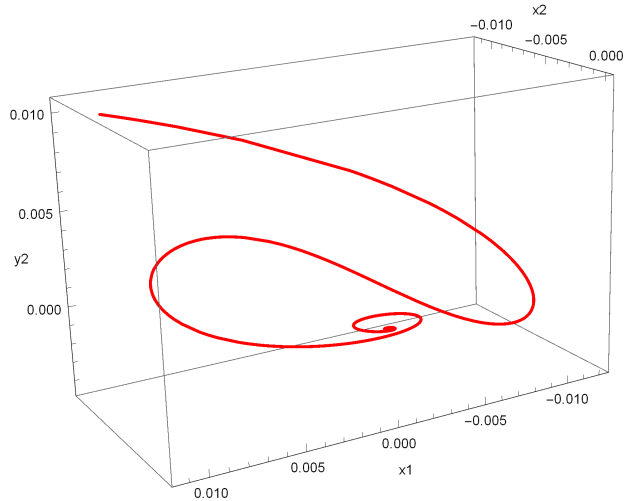}}
\caption{\label{fig:vdp2-a0-08-phase} The solution in $(x_1,x_2,y_2)$ phase space for $a=0.08$ approaching the stable origin from initial conditions.}
\end{figure}

Next, figure \ref{fig:vdp2-a0-08-x1full} shows the solution for $x_1$ for $a =0.08$ after the Hopf Bifurcation, where the origin is now stable. Figure \ref{fig:vdp2-a0-08-phase} shows the solution in $(x_1,x_2,y_2)$ phase space spiraling towards the stable origin. Here we note that the coupled system without delay is still in oscillation (as it does not depend on the delay parameter $a$), that is the delay causes amplitude death where the cyclic coupling alone cannot.   

\subsection{Delayed Sprott System}
Next we consider the numerical results for the delayed Sprott system with cyclic coupling. 

\subsubsection{Trivial Fixed Point}

Here we will consider the case where $\omega = 5$, $\varepsilon = 0.4$. First we note that for this set of parameters each Sprott oscillator is chaotic in isolation (that is uncoupled and without delay), while the coupled system (without delay) is in a state of amplitude death (that is the cyclic coupling results in the trivial fixed point being stable). Also, this parameter set is prior to the transcritical bifurcation at \eqref{transcritical}, and hence fixed point $P_0$ is stable. For this set of parameters, the Routh-Hurwitz conditions at the trivial fixed point of the delayed and coupled system reveal that it Hopf bifurcates at $a\approx8.03562$ and $a\approx 0.773509$. The Routh-Hurwitz conditions at the nontrivial fixed point show us that, for this choice of $(\omega_1,\omega_2,\varepsilon)$, it does not bifurcate as we vary $a$.

\begin{figure}[H]
\centering
\centerline{\includegraphics[width=0.6\textwidth]{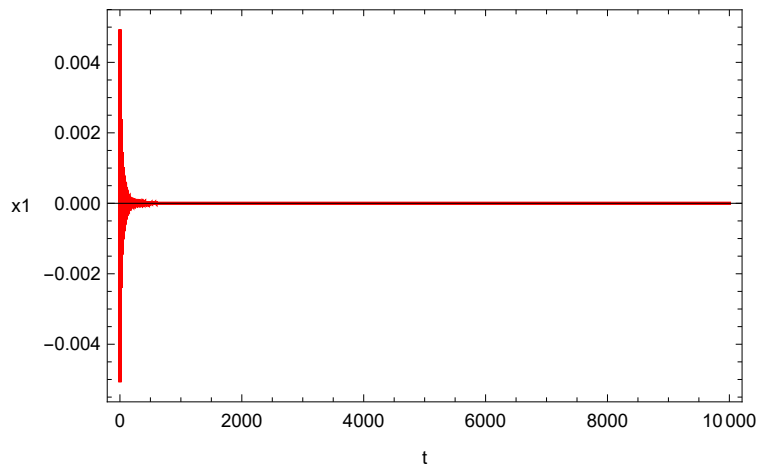}}
\caption{\label{fig:sprott-a14-x1}Amplitude Death in $x_1$ for $a = 14$.}
\end{figure}

\begin{figure}
\centering
\centerline{\includegraphics[width=0.6\textwidth]{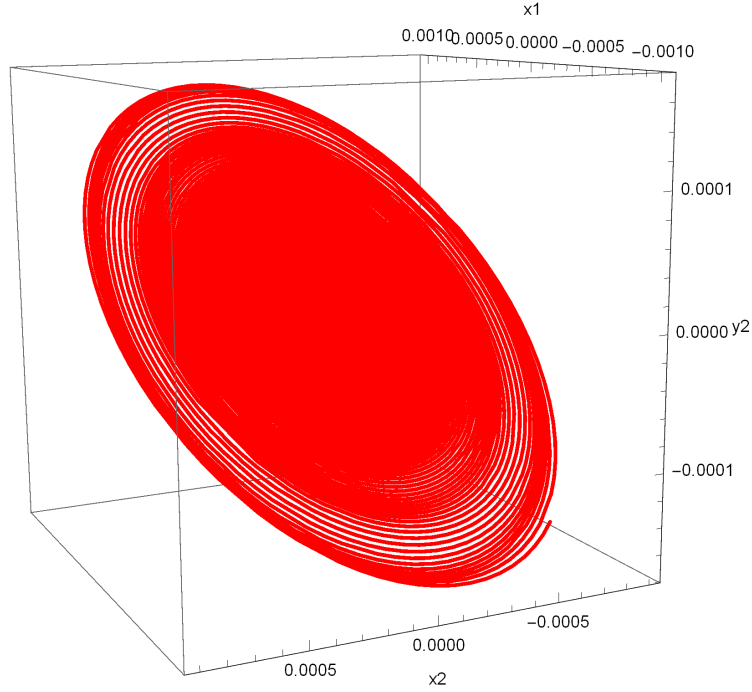}}
\caption{\label{fig:sprott-a14-phase} The solution for $a=14$ spiraling in towards the stable origin from the initial conditions.}
\end{figure}

Figure \ref{fig:sprott-a14-x1}  shows the solution for $x_1$ for $a = 14$ above the first Hopf bifurcation value $a\approx8.03562$. Here, the origin is stable which means we have amplitude death above the first bifurcation point. Figure \ref{fig:sprott-a14-phase} shows the solution in $(x_1,x_2,y_2)$ phase space as the solution spirals towards the origin. 

\begin{figure}
\centering
\centerline{\includegraphics[width=0.6\textwidth]{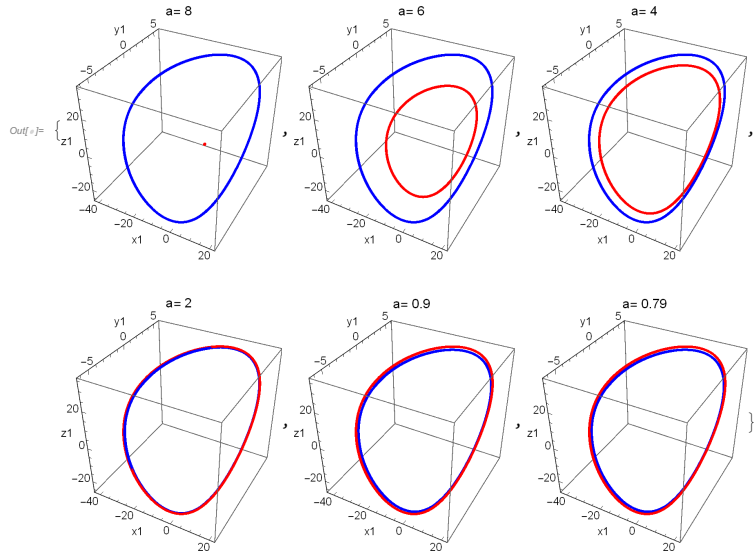}}
\caption{\label{fig:sprott-compare} Phase plane plots for the undelayed Sprott system in blue and the delayed system in red for various values of $a$ between the two bifurcation points.  }
\end{figure}

In figure \ref{fig:sprott-compare} we have plotted the attractor of the isolated (undelayed) Sprott system in blue and the solutions of the delayed, coupled system in red for values of $a = 8,6,4,2,0.9,0.79$ between the two bifurcation points of our system. We observer, as expected, that on this side of the  bifurcation the origin has gone unstable. Here, just below the first bifurcation point at $a=8$, the periodic attractor for the delayed system is very small and close to the origin, and, as we decrease the delay parameter $a$, this limit cycle grows in size towards the attractor of the undelayed isolated system.  

\begin{figure}
\centering
\centerline{\includegraphics[width=0.6\textwidth]{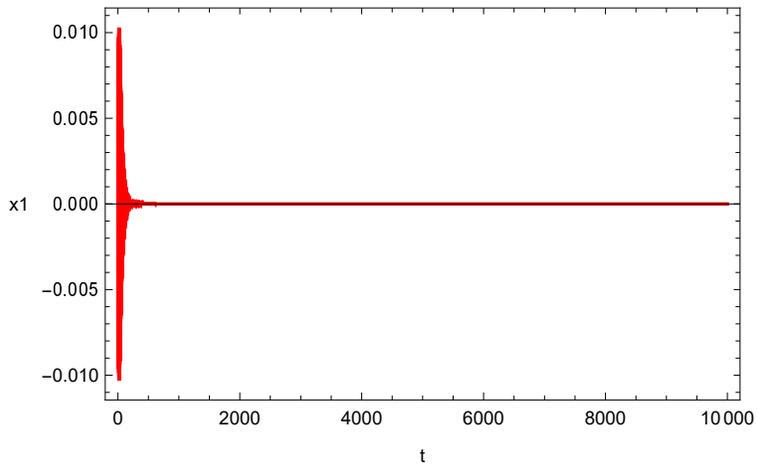}}
\caption{\label{fig:sprott-a0-02-x1} Amplitude Death in $x_1$ for $a = 0.02$. }
\end{figure}

\begin{figure}
\centering
\centerline{\includegraphics[width=0.5\textwidth]{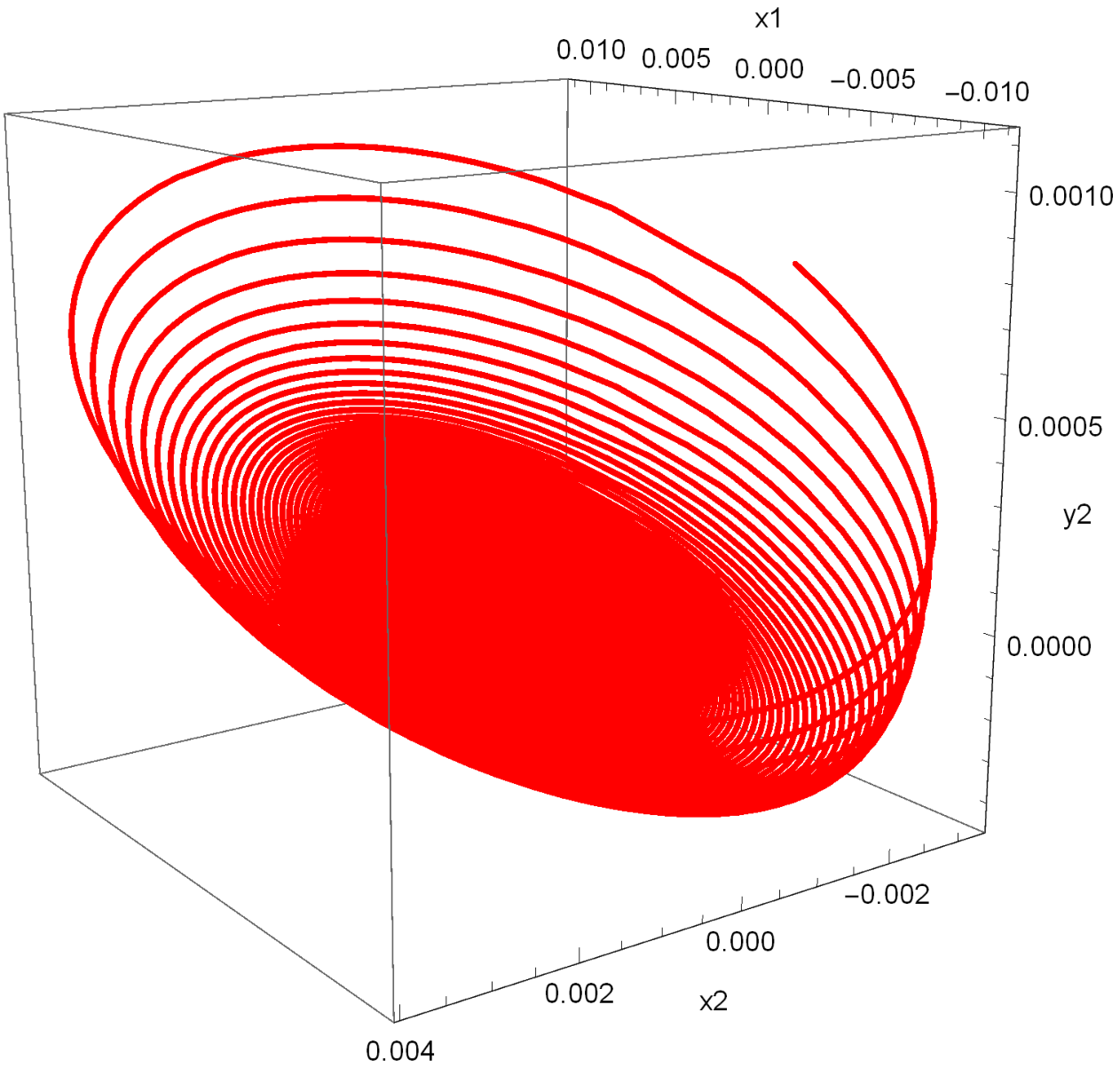}}
\caption{\label{fig:sprott-a0-02-phase} The solution in $(x_1,x_2,y_1)$ parameter space for $a = 0.02$ spiraling towards the stable origin from the initial conditions. }
\end{figure}

Next, Figure \ref{fig:sprott-a0-02-x1} shows the delayed solution for $a=0.02$ below the second bifurcation value $a\approx 0.773509$. Below this second Hopf bifurcation point the delayed system experiences amplitude death as the origin regains stability. Figure \ref{fig:sprott-a0-02-phase} shows the solution in $(x_1,x_2,y_1)$ parameter space approaching the now stable origin from the initial conditions. 

\subsubsection{Nontrivial Fixed Point}

Moving on, we consider the case where $\omega = 1$ and $\varepsilon = 1.41$. This is past the transcritical bifurcation curve \eqref{transcritical}, and hence the trivial fixed point $P_0$ in now unstable. For this set of parameters the Routh Hurwitz conditions at the nontrivial fixed point $P_1$ along with the condition \eqref{d-nontrivx1cond} on for the fixed points, gives us that $x_1^*\approx0.722446$ and the fixed point $P_1$ Hopf bifurcates at $a\approx 0.241658$ and $a\approx 23.8302$. The Routh Hurwitz conditions at the trivial fixed point shows that, for our choice of $(\omega,\varepsilon)$, $P_0$ does not bifurcate as we vary $a$. For these parameters the nontrivial fixed points are given by:
\begin{align}
    P_1 &\approx (0.722446, 0.722446, -2.40815, -1.35563, -1.35563, -1.83772, -2.40815)
\end{align}

\begin{figure}
\centering
\centerline{\includegraphics[width=0.65\textwidth]{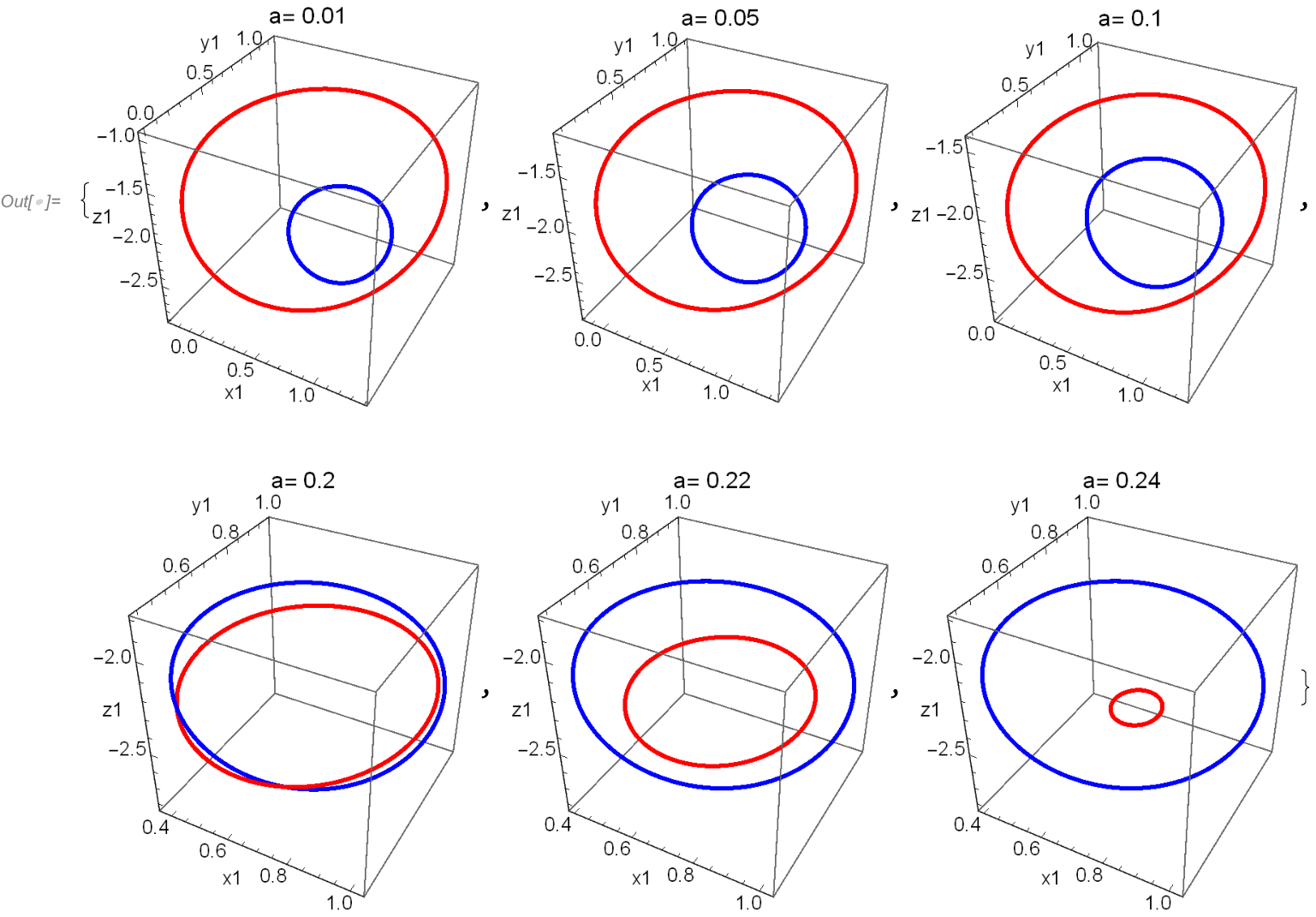}}
\caption{\label{fig:sprottn-compare1}Solutions in $(x_1,y_1,z_1)$ phase space of the coupled system (without delay) in blue and the delayed system in red for $a = 0.01, 0.05,0.1,0.2,0.22,0.24$, before the first bifurcation.}
\end{figure}

\begin{figure}
\centering
\centerline{\includegraphics[width=0.55\textwidth]{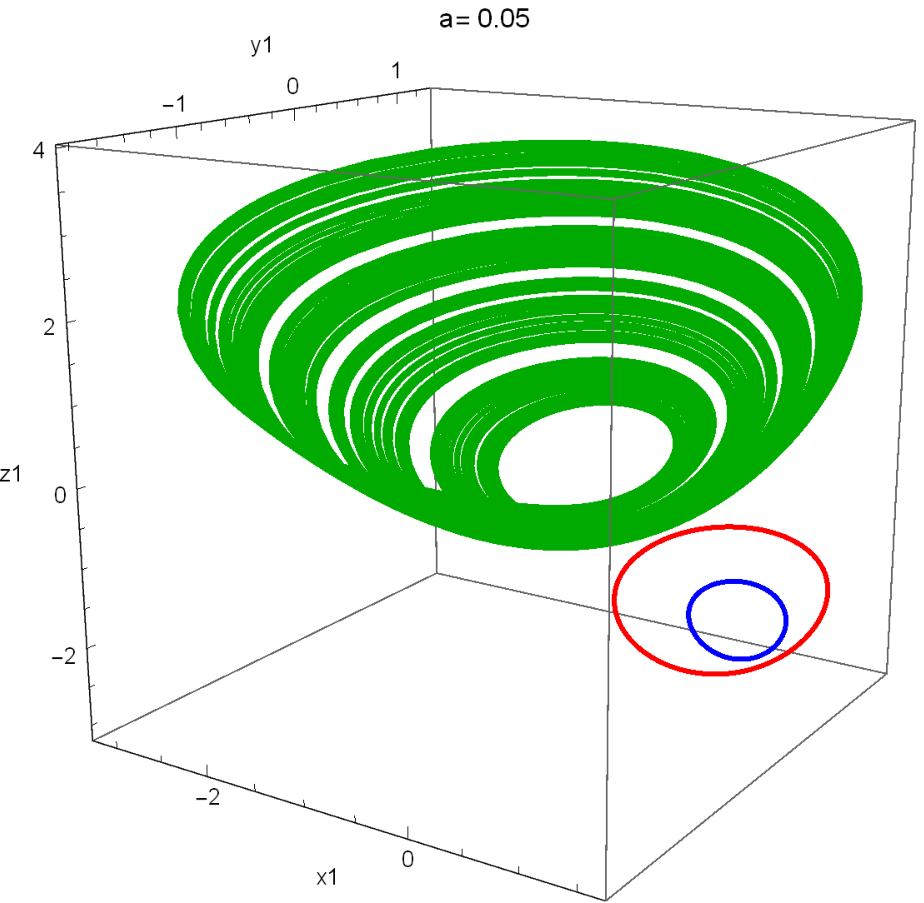}}
\caption{\label{fig:sprottn-compare1-withisolated}Solutions in $(x_1,y_1,z_1)$ phase space of the attractor for the isolated Sprott system in green (without coupling or delay), the solutions for the coupled system without delay in blue and the solution for the delayed system in red for $a =0.05$}
\end{figure}

Figure \ref{fig:sprottn-compare1} shows the solutions in $(x_1,y_1,z_1)$ phase space of the coupled system (without delay) in blue and the delayed system in red for $a = 0.01, 0.05, 0.1,0.2,0.22,0.24$, before the first bifurcation. Here we see that for small values of $a$ the delayed solutions is larger than the undelayed, coupled solution and, as we increase $a$ towards the first bifurcation point $a\approx 0.241658$, the delayed solution shrinks in size around the nontrivial fixed point $P_1$. In figure \ref{fig:sprottn-compare1-withisolated}, we have the attractor for the isolated Sprott system in green (without coupling or delay), the solutions for the coupled system without delay in blue and the solution for the delayed system in red. Here we see that the effects of both the coupling and delay simplifies the behavior of the system. Note that as we further increase $a$ past the first bifurcation point, the delay combined with coupling does what that coupling alone cannot for our parameters, and produces oscillation death:

\begin{figure}
\centering
\centerline{\includegraphics[width=0.6\textwidth]{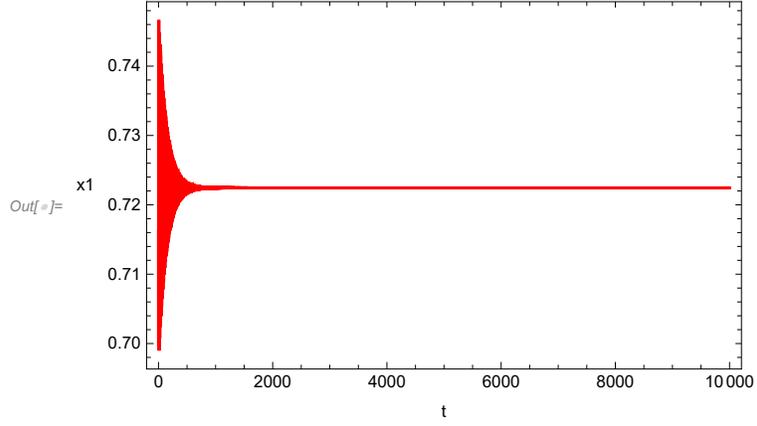}}
\caption{\label{fig:sprottn-a12-x1} Oscillation death in $x_1$ for $a = 12$}
\end{figure}

\begin{figure}
\centering
\centerline{\includegraphics[width=0.5\textwidth]{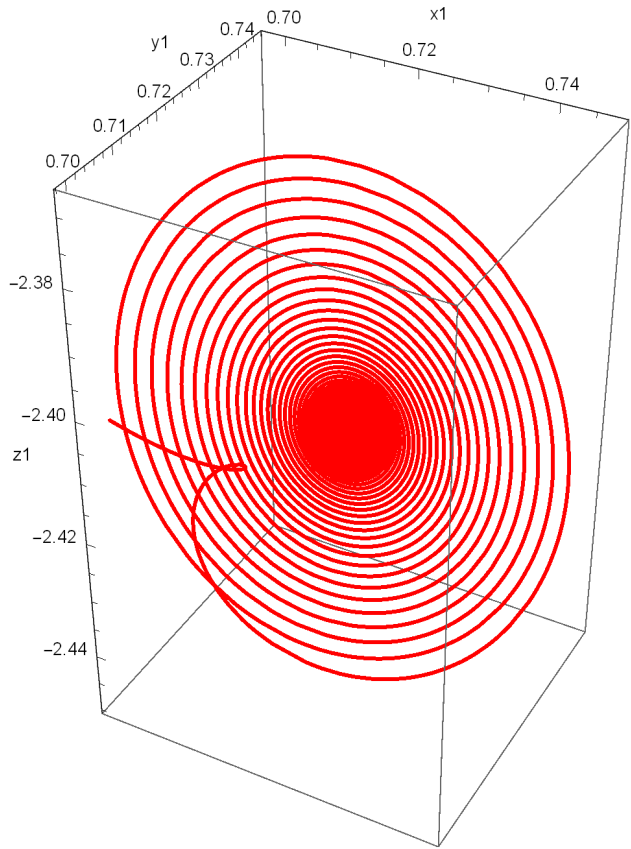}}
\caption{\label{fig:sprottn-a12-phase} Solution in $(x_1,y_1,z_1)$ phase space for $a=12$ and approach to the fixed point $P_1$ from initial conditions.}
\end{figure}

After the first Hopf bifurcation at $a\approx 0.241658$, the nontrivial fixed point becomes stable. In figure \ref{fig:sprottn-a12-x1} we have the solution for the delayed system for $a=12$ showing the $x_1$ solution approaching the fixed point. Figure \ref{fig:sprottn-a12-phase} shows the solution in $(x_1,y_1,z_1)$ phase space and the approach towards the fixed point $P_1$ from the initial conditions.

\begin{figure}
\centering
\centerline{\includegraphics[width=0.6\textwidth]{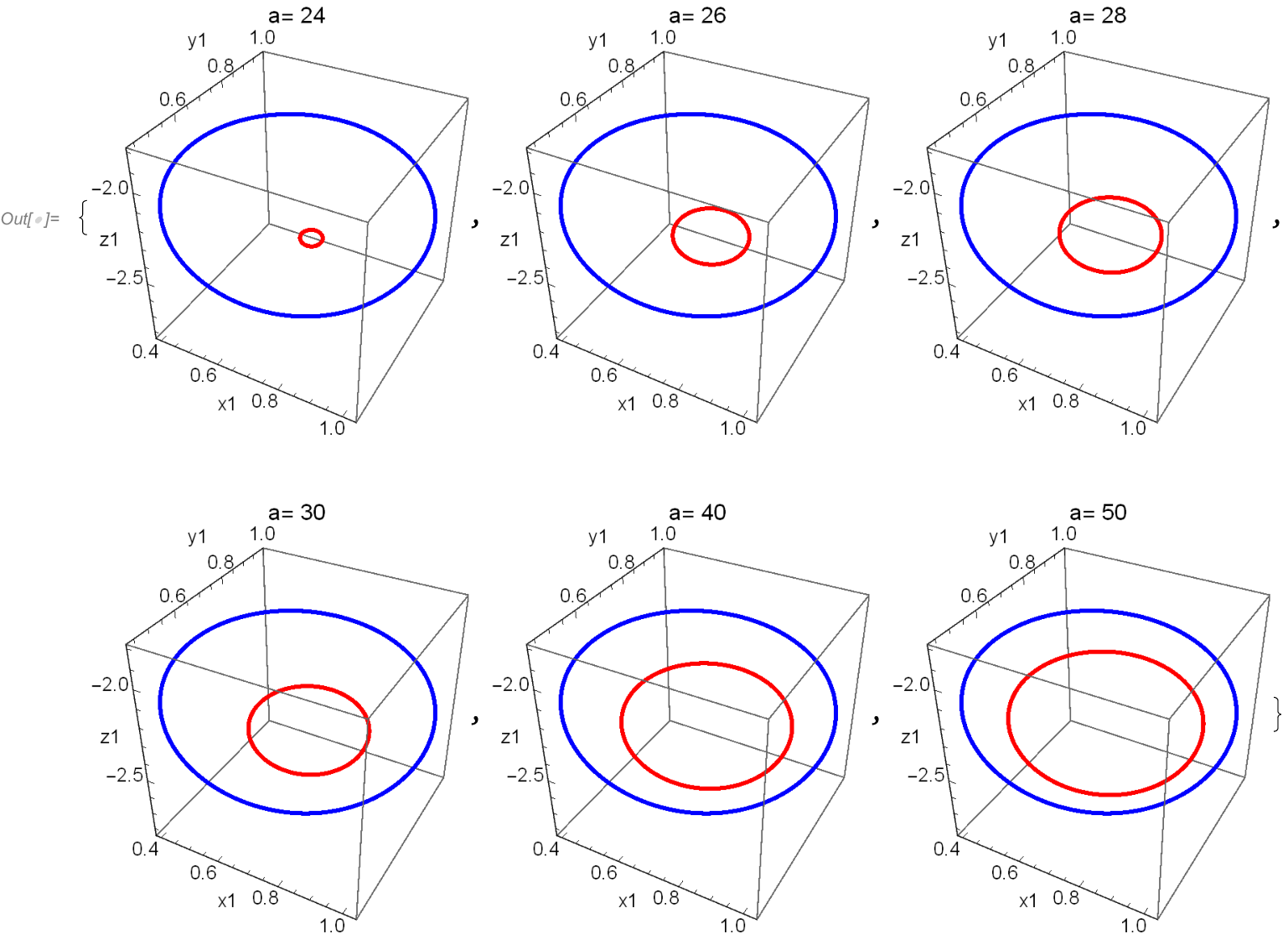}}
\caption{\label{fig:sprottn-compare2} Solutions of the coupled, undelayed system in blue and the delayed system in red for values of $a=24,26,28,30,40,50$ in $(x_1,y_1,z_1)$ phase space.}
\end{figure}

\begin{figure}[H]
\centering
\centerline{\includegraphics[width=0.5\textwidth]{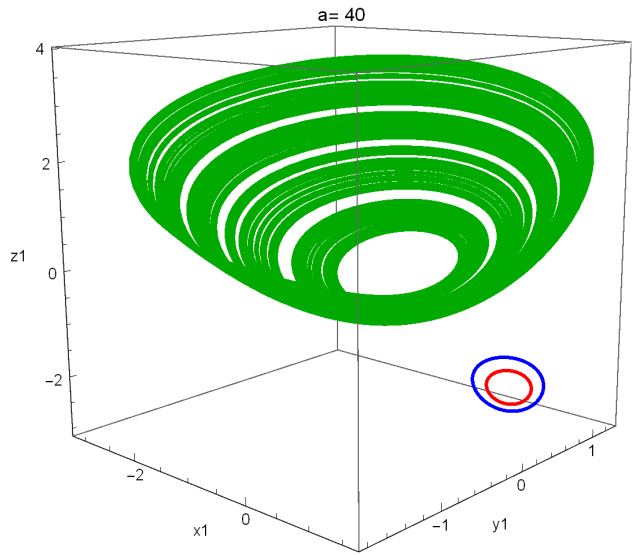}}
\caption{\label{fig:sprottn-compare2-withisolated}Solutions in $(x_1,y_1,z_1)$ phase space of the attractor for the isolated Sprott system in green (without coupling or delay), the solutions for the coupled system without delay in blue and the solution for the delayed system in red for $a =40$}
\end{figure}

Upon further increasing the delay parameter $a$ past the second bifurcation value $a\approx 23.8302$ we find that the nontrivial fixed point loses its stability. Figure \ref{fig:sprottn-compare2} shows the solution of the coupled, undelayed system in blue and the delayed system in red for several values of $a$ past the second bifurcation point in $(x_1,y_1,z_1)$ phase space. Here we see that initially, after the bifurcation, the delayed periodic solution is very small, still orbiting close to the fixed point and as we increase the value for $a$ the orbit for the delayed solution grows in size. Figure \ref{fig:sprottn-compare2-withisolated} shows the periodic solutions in $(x_1,y_1,z_1)$ phase space for the isolated Sprott system in green (without coupling or delay), the solutions for the coupled system without delay in blue, and the solution for the delayed system in red for $a =40$. Once again, we see that the delay plus coupling has simplified the motion of the system.  

\section{Discussion and Conclusions}

In this paper, we have systematically considered the effects of a distributed 'weak generic kernel' exponential delay on both cyclically coupled limit cycle and chaotic oscillators. The effects of the delay are similar, both for coupled Van der Pol oscillators and in fact, other oscillators as well, where the delay can produce transitions from AD/OD to periodic orbits via Hopf bifurcation, with the delayed limit cycle shrinking or growing as we vary the delay towards or away from the bifurcation point respectively \cite{Paper1}-\cite{Paper2}. The transition from AD to OD is mediated here via a pitchfork bifurcation, as seen earlier for other couplings as well \cite{Suz, Sha}. Also, the cyclically coupled van der Pol system here is already in a state of AD/OD, and introducing the delay allows both oscillations and AD/OD as the delay parameter is varied. This is in contrast to \cite{Paper1} for example, where the diffusive coupling alone did not result in the onset of AD/OD. 

For systems whose isolated systems are chaotic, such as the Sprott system in this paper, or a coupled van der Pol-Rayleigh system with parametric forcing\cite{Paper1}), we see that the delay {\it may} produce AD/OD (as in the Sprott case), with the AD to OD transition now however mediated by a transcritical bifurcation. However, this might not be possible, and the delay might just vary the attractor shape\cite{Paper1}. In both cases however,  we see that increased delay strength tends to cause the system to have simpler behavior, simplifying the shape of the attractor, or shrinking it in cases with limit cycle behavior.

\appendix
\section{Coefficients in characteristic equation \eqref{characnew}}\label{AppA}
\setcounter{equation}{0}

The coefficients in \eqref{characnew} are:
\begin{align}
    b_1 &= \frac{1}{10 \varepsilon \omega_2 \left(\varepsilon^2+\omega_1 \omega_2\right)^2}\bigg(10 a \varepsilon^5 \omega_2+2 (10 a-3) \varepsilon^3 \omega_1 \omega_2^2+(10 a-3) \varepsilon \omega_1^2 \omega_2^3\notag\\
    &\quad-20 \varepsilon^6 (\omega_1-\omega_2)-40 \varepsilon^4 \omega_1 \omega_2 (\omega_1-\omega_2)+10 \varepsilon^2 \omega_1^2 \omega_2^2 (2 \omega_2-3 \omega_1)-10 \omega_1^4 \omega_2^3\bigg)\\
    b_2 &= \frac{1}{10 \varepsilon \omega_2^2 \left(\varepsilon^2+\omega_1 \omega_2\right)^2}\bigg(-20 a \varepsilon^6 \omega_2 (\omega_1-\omega_2)+2 \varepsilon^4 \omega_1 \omega_2^2 ((3-20 a) \omega_1+20 a \omega_2)\notag\\
    &\quad+2 \varepsilon^3 \omega_1 \omega_2^2 \left(-3 a \omega_2+5 \omega_1^3-40 \omega_1^2 \omega_2+5 \omega_1 \omega_2^2+10 \omega_2^3\right)+\varepsilon^2 \omega_1^2 \omega_2^3 ((9-30 a) \omega_1\notag\\
    &\quad+(20 a+3) \omega_2)+\varepsilon \omega_1^2 \omega_2^4 \left(10 \left(\omega_2^2-3 \omega_1^2\right)-3 a\right)+(3-10 a) \omega_1^4 \omega_2^4\notag\\
    &\quad+10 \varepsilon^7 \left(\omega_1^2-3 \omega_1 \omega_2+\omega_2^2\right)+10 \varepsilon^5 \omega_2 \left(2 \omega_1^3-8 \omega_1^2 \omega_2+2 \omega_1 \omega_2^2+\omega_2^3\right)\bigg)\\
    b_3 &= \frac{1}{50 \varepsilon \omega_2^2 \left(\varepsilon^2+\omega_1 \omega_2\right)^3}\bigg( 5 a \left(10 \varepsilon^9 \left(\omega_1^2-3 \omega_1 \omega_2+\omega_2^2\right)+10 \varepsilon^7 \omega_2 \left(3 \omega_1^3-11 \omega_1^2 \omega_2\right.\right.\notag\\
    &\quad\left. +3 \omega_1 \omega_2^2+\omega_2^3\right)+6 \varepsilon^6 \omega_1^2 \omega_2^2+10 \varepsilon^5 \omega_1 \omega_2^2 \left(3 \omega_1^3-16 \omega_1^2 \omega_2+3 \omega_1 \omega_2^2+3 \omega_2^3\right)\notag\\
    &\quad+3 \varepsilon^4 \omega_1^2 \omega_2^3 (5 \omega_1+\omega_2)+10 \varepsilon^3 \omega_1^2 \omega_2^3 \left(\omega_1^3-11 \omega_1^2 \omega_2+\omega_1 \omega_2^2+3 \omega_2^3\right)\notag\\
    &\quad+3\left. \varepsilon^2 \omega_1^3 \omega_2^4 (4 \omega_1+\omega_2)+10 \varepsilon \omega_1^3 \omega_2^5 \left(\omega_2^2-3 \omega_1^2\right)+3 \omega_1^5 \omega_2^5\right)+50 \varepsilon^{10} \omega_1 (\omega_1-\omega_2)\notag\\
    &\quad+50 \varepsilon^8 \omega_2 \left(4 \omega_1^3-5 \omega_1^2 \omega_2-\omega_1 \omega_2^2+\omega_2^3\right)+30 \varepsilon^7 \omega_1 \omega_2^3+50 \varepsilon^6 \omega_1 \omega_2^2 \left(5 \omega_1^3-9 \omega_1^2 \omega_2\right.\notag\\
    &\quad\left. -4 \omega_1 \omega_2^2+3 \omega_2^3\right)+15 \varepsilon^5 \omega_1^2 \omega_2^3 (5 \omega_1+4 \omega_2)+2 \varepsilon^4 \omega_1^2 \omega_2^3 \left(50 \omega_1^3-175 \omega_1^2 \omega_2\right.\notag\\
    &\quad\left. -150 \omega_1 \omega_2^2+75 \omega_2^3-9 \omega_2\right)+15 \varepsilon^3 \omega_1^3 \omega_2^4 (7 \omega_1+2 \omega_2)-\varepsilon^2 \omega_1^3 \omega_2^5 \left(100 \omega_1^2\right.\notag\\
    &\quad\left. +200 \omega_1 \omega_2-50 \omega_2^2+9\right)+30 \varepsilon \omega_1^5 \omega_2^5-50 \omega_1^5 \omega_2^7\bigg)\\
    b_4 &= \frac{a}{50 \varepsilon \omega_2^2 \left(\varepsilon^2+\omega_1 \omega_2\right)^3}\bigg(50 \varepsilon^{10} \omega_1 (\omega_1-\omega_2)+50 \varepsilon^8 \omega_2 \left(4 \omega_1^3-5 \omega_1^2 \omega_2-\omega_1 \omega_2^2+\omega_2^3\right)\notag\\
    &\quad+30 \varepsilon^7 \omega_1 \omega_2^3+50 \varepsilon^6 \omega_1 \omega_2^2 \left(5 \omega_1^3-9 \omega_1^2 \omega_2-4 \omega_1 \omega_2^2+3 \omega_2^3\right)+15 \varepsilon^5 \omega_1^2 \omega_2^3 (5 \omega_1+4 \omega_2)\notag\\
    &\quad+2 \varepsilon^4 \omega_1^2 \omega_2^3 \left(50 \omega_1^3-175 \omega_1^2 \omega_2-150 \omega_1 \omega_2^2+75 \omega_2^3-9 \omega_2\right)+15 \varepsilon^3 \omega_1^3 \omega_2^4 (7 \omega_1+2 \omega_2)\notag\\
    &\quad-\varepsilon^2 \omega_1^3 \omega_2^5 \left(100 \omega_1^2+200 \omega_1 \omega_2-50 \omega_2^2+9\right)+30 \varepsilon \omega_1^5 \omega_2^5-50 \omega_1^5 \omega_2^7\bigg)\notag\\
    &\quad-\frac{\omega_1 \omega_2 \left(5 \varepsilon^4+15 \varepsilon^2 \omega_1 \omega_2-3 \varepsilon \omega_2^2+10 \omega_1^2 \omega_2^2\right)}{5 \left(\varepsilon^2+\omega_1 \omega_2\right)}\\
    b_5 &= \frac{1}{5} a \omega_2 \left(-10 \varepsilon^2 \omega_1+3 \varepsilon \omega_2-10 \omega_1^2 \omega_2\right)
\end{align}
where
\begin{align}
    c_1 &=  \frac{\varepsilon^2 \omega_1-\frac{3 \varepsilon \omega_2}{5}+\omega_1^2 \omega_2}{\varepsilon^2+\omega_1 \omega_2}\\
    c_2 &= -\frac{3 \varepsilon^4}{10 \left(\varepsilon^2+\omega_1 \omega_2\right)^2}+\frac{\varepsilon^3 \omega_1}{\varepsilon^2 \omega_2+\omega_1 \omega_2^2}-\varepsilon+\frac{3}{10}
\end{align}

\section{Coefficients in characteristic equations \eqref{efunc2} and \eqref{efunc2n}}\label{AppB}
\setcounter{equation}{0}

The coefficients in  \eqref{efunc2} are:
\begin{align}
b_1 &= a+2 \varepsilon+\frac{7}{5} \notag \\
b_2 &= a \left(2 \varepsilon+\frac{7}{5}\right)+\varepsilon^2+\frac{31 \varepsilon}{10}+2 \omega^2-\frac{11}{100} \notag \\
b_3 &= \frac{1}{100} \left(100 a \varepsilon^2+310 a \varepsilon+200 a \omega^2-11 a+170 \varepsilon^2+200 \varepsilon \omega^2+29 \varepsilon\right.\\
&\quad\quad\quad\quad\left.+340 \omega^2-42\right) \notag \\
b_4 &= \frac{1}{100} \left(170 a \varepsilon^2+200 a \varepsilon \omega^2+29 a \varepsilon+340 a \omega^2-42 a+40 \varepsilon^2+370 \varepsilon \omega^2\right.\notag\\
&\quad\quad\quad\quad\left.-72 \varepsilon+100 \omega^4+80 \omega^2+9\right) \notag \\
b_5 &= \frac{1}{100} \left(-100 a \varepsilon^2 \omega^2+40 a \varepsilon^2+370 a \varepsilon \omega^2-72 a \varepsilon+100 a \omega^4+80 a \omega^2+9 a\right.\notag\\
&\quad\quad\quad\quad\left.-30 \varepsilon^2+140 \varepsilon \omega^2+9 \varepsilon+200 \omega^4-60 \omega^2\right)\notag \\
b_6 &= \frac{1}{100} \left(-200 a \varepsilon^2 \omega^2-30 a \varepsilon^2+140 a \varepsilon \omega^2+9 a \varepsilon+200 a \omega^4-60 a \omega^2\right.\notag\\
&\quad\quad\quad\quad\left.-30 \varepsilon \omega^2+100 \omega^4\right)\notag \\
b_7 &= \frac{1}{10} a \omega^2 \left(-10 \varepsilon^2-3 \varepsilon+10 \omega^2\right)
\end{align}

And the coefficients in \eqref{efunc2n} are:
\begin{align}
b_1 &= a+2 \varepsilon-x_1^*-x_2^*+\frac{7}{5} \notag \\
b_2 &= a \left(2 \varepsilon-x_1^*-x_2^*+\frac{7}{5}\right)+\varepsilon^2+\varepsilon \left(-x_1^*-2 x_2^*+\frac{31}{10}\right)+2 \omega^2+x_1^* x_2^*\notag\\
&\quad-\frac{12 x_1^*}{5}-\frac{12 x_2^*}{5}-\frac{11}{100} \notag \\
b_3 &= \frac{1}{100} \left(a \left(100 \varepsilon^2-10 \varepsilon (10 x_1^*+20 x_2^*-31)+200 \omega^2+100 x_1^* x_2^*-240 x_1^*\right.\right.\notag\\
&\quad \left.\left.-240 x_2^*-11\right)+\varepsilon^2 (170-100 x_2^*)+\varepsilon \left(200 \omega^2+100 x_1^* x_2^*-270 x_1^*-510 x_2^*\right.\right.\notag\\
&\quad \left.\left.+29\right)-100 \omega^2 x_1^*-100 \omega^2 x_2^*+340 \omega^2+340 x_1^* x_2^*-29 x_1^*-29 x_2^*-42\right) \notag \\
b_4 &= \frac{1}{100} \left(-a \left(10 \varepsilon^2 (10 x_2^*-17)+\varepsilon \left(-200 \omega^2-100 x_1^* x_2^*+270 x_1^*+510 x_2^*-29\right)\right.\right.\notag\\
&\quad \left.\left.+20 \omega^2 (5 x_1^*+5 x_2^*-17)-340 x_1^* x_2^*+29 x_1^*+29 x_2^*+42\right)+\varepsilon^2 (40-270 x_2^*)\right.\notag\\
&\quad \left.-\varepsilon \left(10 \omega^2 (10 x_2^*-37)+x_1^* (110-370 x_2^*)+139 x_2^*+72\right)+100 \omega^4\right.\notag\\
&\quad \left.-270 \omega^2 x_1^*-270 \omega^2 x_2^*+80 \omega^2+169 x_1^* x_2^*+93 x_1^*+93 x_2^*+9\right) \notag \\
b_5 &= \frac{1}{100} \left(a \left(-10 \varepsilon^2 \left(10 \omega^2+27 x_2^*-4\right)-\varepsilon \left(10 \omega^2 (10 x_2^*-37)-370 x_1^* x_2^*+110 x_1^*\right.\right.\right.\notag\\
&\quad \left.\left.\left.+139 x_2^*+72\right)+100 \omega^4-10 \omega^2 (27 x_1^*+27 x_2^*-8)+169 x_1^* x_2^*+93 x_1^*+93 x_2^*+9\right)\right.\notag\\
&\quad \left.-10 \varepsilon^2 (11 x_2^*+3)+\varepsilon \left(-20 \omega^2 (15 x_2^*-7)+20 x_1^* (14 x_2^*+3)+153 x_2^*+9\right)+200 \omega^4\right.\notag\\
&\quad \left.-10 \omega^2 (11 x_1^*+11 x_2^*+6)-6 (x_1^* (34 x_2^*+3)+3 x_2^*)\right) \notag \\
b_6 &= \frac{1}{100} \left(2 \left(3 \varepsilon-10 \omega^2-6 x_1^*\right) \left((10 \varepsilon-3) x_2^*-5 \omega^2\right)-a \left(10 \varepsilon^2 \left(20 \omega^2+11 x_2^*+3\right)\right.\right.\notag\\
&\quad \left.\left.+\varepsilon \left(20 \omega^2 (15 x_2^*-7)-280 x_1^* x_2^*-60 x_1^*-153 x_2^*-9\right)+2 \left(-100 \omega^4+5 \omega^2 (11 x_1^*\right.\right.\right.\notag\\
&\quad \left.\left.\left.+11 x_2^*+6)+102 x_1^* x_2^*+9 x_1^*+9 x_2^*\right)\right)\right) \notag \\
b_7 &= \frac{1}{50} a \left(\varepsilon^2 \left(30 x_2^*-50 \omega^2\right)-5 \varepsilon \omega^2 (20 x_2^*+3)-3 \varepsilon (20 x_1^*+3) x_2^*\right.\notag\\
&\quad \left.+2 \left(5 \omega^2+3 x_1^*\right) \left(5 \omega^2+3 x_2^*\right)\right)
\end{align}

\section{Summary Statement}

Conflict of Interest: The authors declare that they have no conflict of interest.

\end{document}